\documentclass{article}
\usepackage{graphicx} 
\usepackage[colorlinks=true, allcolors=blue]{hyperref}
\usepackage{makecell}
\usepackage[table,xcdraw]{xcolor}
\usepackage{subfigure}
\usepackage[super]{nth}
\usepackage{float}
\usepackage{xurl}
\usepackage{multirow}
\usepackage{multicol}
\usepackage{adjustbox}
\usepackage{pifont}
\usepackage{float}
\usepackage[most]{tcolorbox}
\tcbuselibrary{breakable}
\usepackage{enumitem}
\usepackage{authblk}
\newtheorem{exmp}{Example}[section]

\title{Securing the Internet of Medical Things (IoMT): Real-World Attack Taxonomy and Practical Security Measures}
\author[$\star$]{Suman Deb}
\author[$\star$]{Emil Lupu}
\author[$\star$]{Emm Mic Drakakis}
\author[$\star$]{Anil Anthony Bharath}
\author[$\ddagger$]{Zhen Kit Leung}
\author[$\ddagger$]{Guang Rui Ma}
\author[$\star$, $\dagger$]{Anupam Chattopadhyay}
\affil[$\dagger$]{College of Computing and Data Science, NTU, Singapore}
\affil[$\star$]{Imperial Global Singapore, Singapore}
\affil[$\ddagger$]{SingHealth BME, Singapore}

\begin{document}
\maketitle

\begin{abstract}
The Internet of Medical Things (IoMT) has the potential to radically improve healthcare by enabling real-time monitoring, remote diagnostics, and AI-driven decision making. However, the connectivity, embedded intelligence, and inclusion of a wide variety of novel sensors expose medical devices to severe cybersecurity threats, compromising patient safety and data privacy. In addition, many devices also have direct capacity \textemdash individually or in conjunction with other IoMT devices \textemdash to perform actions on the patient, such as delivering an electrical stimulus, administering a drug, or activating a motor, which can potentially be life-threatening. We provide a taxonomy of potential attacks targeting IoMT, presenting attack surfaces, vulnerabilities, and mitigation strategies across all layers of the IoMT architecture. It answers key questions such as: What makes IoMT security different from traditional IT security? What are the cybersecurity threats to medical devices? How can engineers design secure IoMT systems and protect hospital networks from cyberattacks? By analyzing historical cyber incidents, we highlight critical security gaps and propose practical security guidelines for medical device engineers and security professionals. This work bridges the gap between research and implementation, equipping healthcare stakeholders with actionable insights to build resilient and privacy-preserving IoMT ecosystems. Finally, we present the latest standardization and compliance frameworks, that IoMT security designers should be aware of.
\end{abstract}

\section{Introduction}
\begin{figure}[H]
    \centering
    \includegraphics[width=\linewidth]{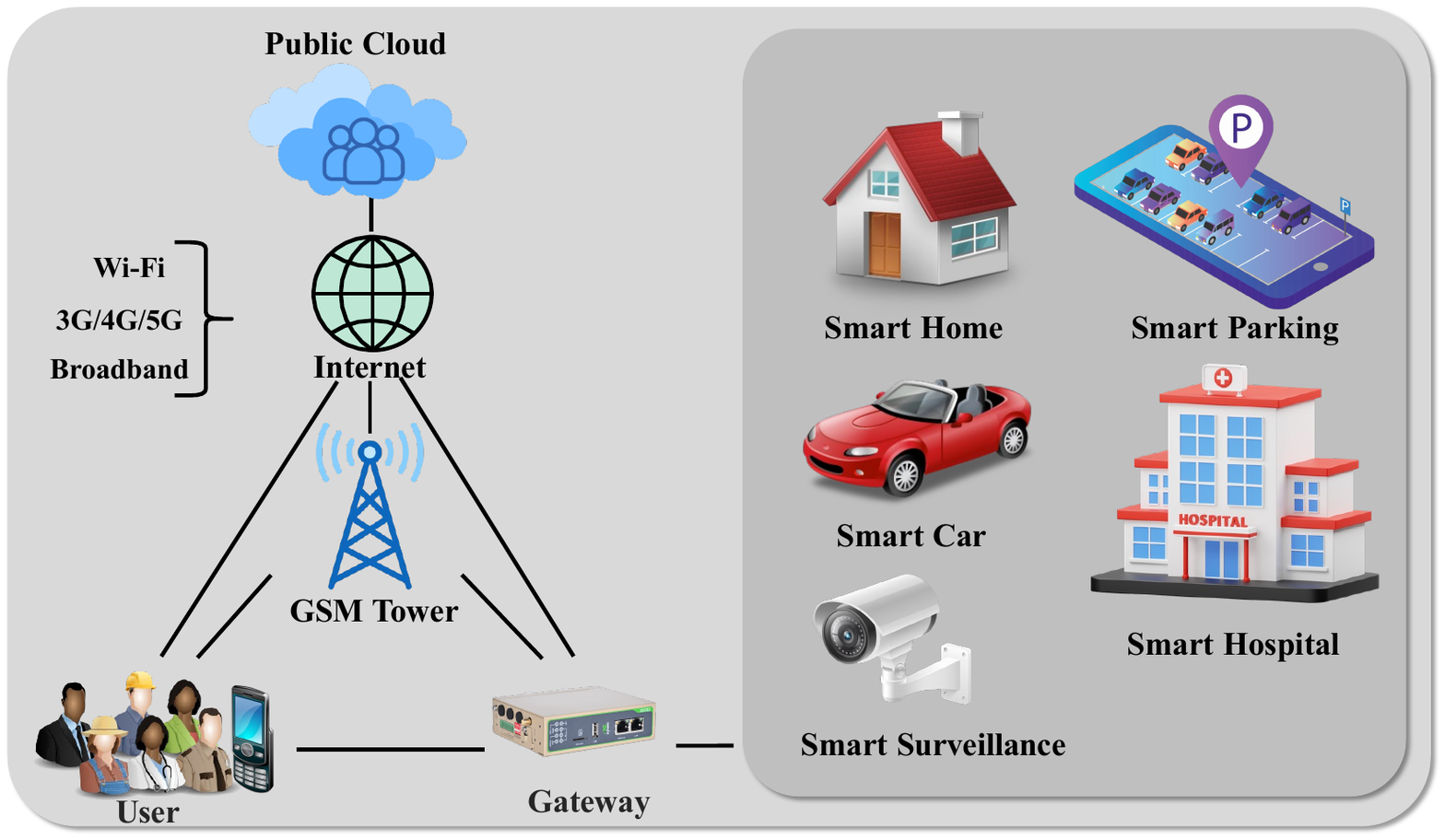}
    \caption{An Overview of Internet of Things (IoT) and its Applications}
    \label{fig:iot}
\end{figure}

Over the last two decades, the internet has rapidly evolved from being a network of primarily desktop computers to a heterogeneous network of diverse electronic objects. This network of interconnected objects is popularly known as the `Internet of Things' (IoT). IoT creates an intelligent network that not only collects (senses) data from the physical world and interacts (actuation) with its environment, but also uses internet standards for efficient transfer, storage, and analysis of data streams. The concept of IoT is shown in Figure~\ref{fig:iot}. Using technologies such as silicon microfabrication, wireless communication, and cloud computing as its building blocks, IoT has grown at such an unprecedented rate that the number of interconnected devices in the world exceeded the total number of people on Earth as early as 2011~\cite{gubbi2013internet}. Aksu \textit{et al.} reported in~\cite{aksu2018identification} that two new devices are connected to the internet every 3 minutes. 

\begin{figure}[H]
    \centering
    \includegraphics[width=\linewidth]{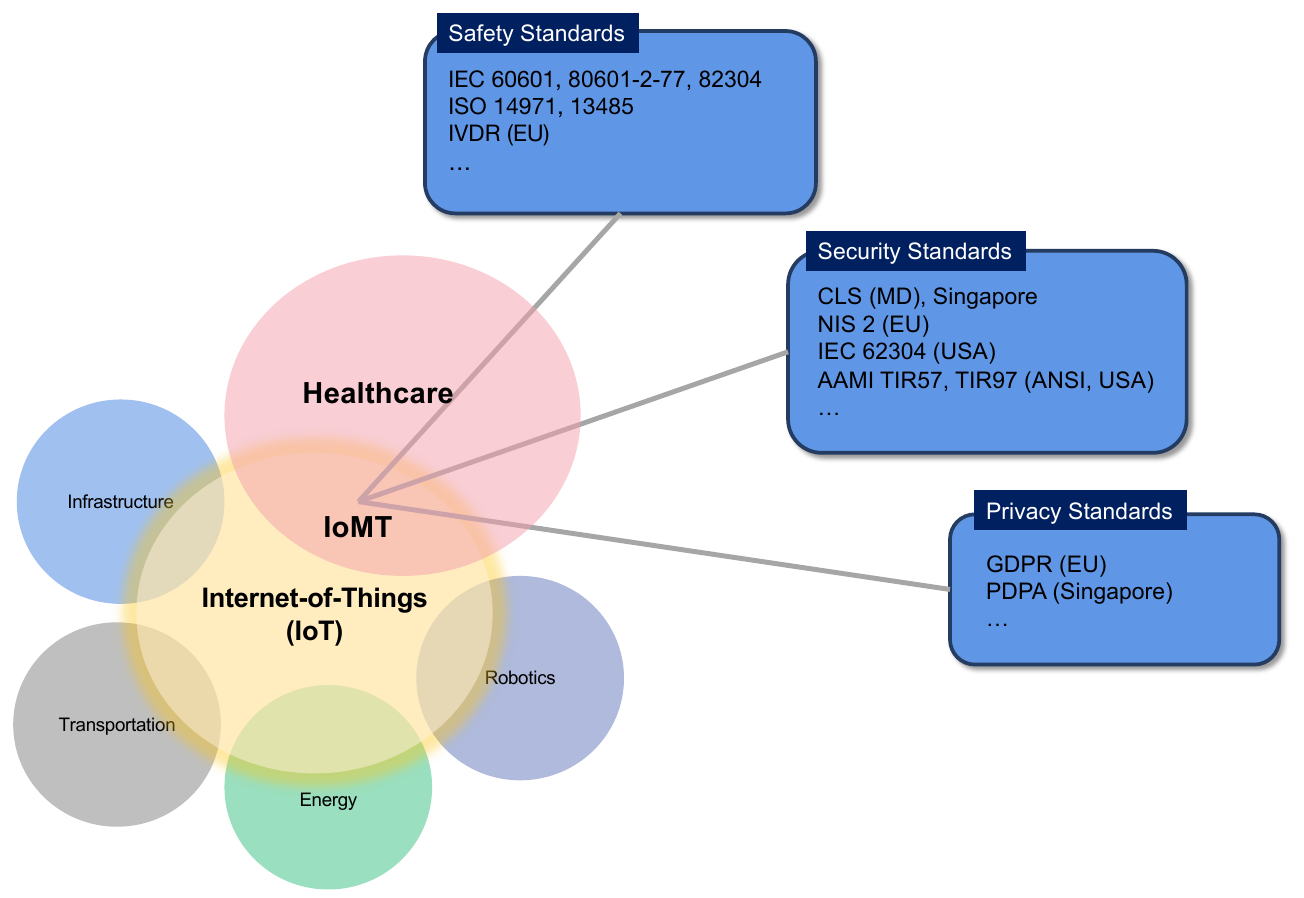}
    \caption{Applications of IoT and Evolving Standards}
    \label{fig:iotvenn}
\end{figure}

As a result, although the term `Internet of Things' was coined by Kevin Ashton in 1999 in the context of supply chain management~\cite{ashton2009internet}, the definition of `Things' now covers a broad range of applications. These applications, as shown in Figure~\ref{fig:iotvenn}, span from healthcare (Internet of Medical Things (IoMT)) to industrial automation (Industrial Internet of Things (IIoT)) to transport (Internet of Vehicles (IoV)). Along the same lines, IoMT is bringing the benefits of digitization, distributed intelligence, and connectivity to healthcare. For example, using wireless RF and Bluetooth, implanted devices such as pacemakers and neuro-stimulators can now be adjusted post-implantation, without further surgery, to refine the management of cardiac arrhythmia. Pacemakers can detect subtle changes in heart rhythm and not only attempt to correct them, but also send data through a personal mobile equipped with an app to the medical clinic for review. Continuous glucose monitors can communicate with insulin pumps, enabling better blood sugar control for Type 1 diabetics. 

For patients with kidney disorders, dialysis can be performed at home and doctors can monitor treatment remotely. Devices can capture and send data about a therapeutic session or intervention episode to clinics and manufacturers for analysis, helping to monitor treatments. Such information can even be used to improve treatment protocols. Electronic Health Records (EHRs) can be adapted to store data
from medical devices to improve care coordination and reduce errors.

\subsection{Background}
According to the Health Sciences Authority (HSA) of Singapore~\cite{hsasg}, a medical device is defined as follows. Any instrument, apparatus, implement, machine, appliance, implant, reagent for in vitro use, software, material, or other similar or related article that is intended by its manufacturer to be used, whether alone or
in combination, for humans for one or more of the specific purposes of the following.
\begin{itemize}
    \item Diagnosis, prevention, monitoring, treatment or alleviation of disease
    \item Diagnosis, monitoring, treatment or alleviation of, or compensation for, an injury
    \item Investigation, replacement, modification or support of the anatomy or of a physiological process, mainly for medical purposes
    \item Supporting or sustaining life
    \item Control of contraception
    \item Disinfection of medical devices
    \item Providing information by means of in vitro examination of specimens derived from the human body, for medical or diagnostic purposes 
\end{itemize}
Furthermore, a medical device does not achieve its primary intended action in or on the human body by pharmacological, immunological, or metabolic means.

Medical devices are distinct from drugs and biologics as they generally achieve their intended purpose primarily through connected, digital means, and treat the end drug as an agent for biological and chemical action. The World Health Organization (WHO) estimates that the current global market comprises approximately 2 million distinct types of medical devices, organized into more than 7,000 generic device categories~\cite{who}. Table~\ref{tab:med_devices} presents some of the common categories of medical devices along with examples.

\begin{table}[H]
\footnotesize
    \centering
    \caption{Common Types of Medical Devices}
    \label{tab:med_devices}
\begin{tabular}{cc}
    \hline
    \rowcolor{gray!30} \textbf{Device Type} & \textbf{Examples of Medical Devices} \\ \hline

    \textbf{Diagnostic} & Blood Glucose Meter, Digital Thermometer, MRI Machine \\ 
    \textbf{Therapeutic} & Insulin Pump, Pacemaker, Dialysis Machine \\ 
    \textbf{Monitoring} & Wearable Heart Monitor, Fetal Monitor, Pulse Oximeter \\ 
    \textbf{Surgical} & Robotic Surgery System, Endoscope \\ 
    \textbf{Home Healthcare} & CPAP Machine, Digital Blood Pressure Monitor, Nebulizer \\ 
    \textbf{Implantable} & Cochlear Implant, Artificial Heart Valve, Spinal Cord Stimulator \\ \hline
    \end{tabular}
\end{table}

\subsection{Architecture of IoMT System}
\begin{figure}[H]
    \centering
    \includegraphics[width=0.96\linewidth]{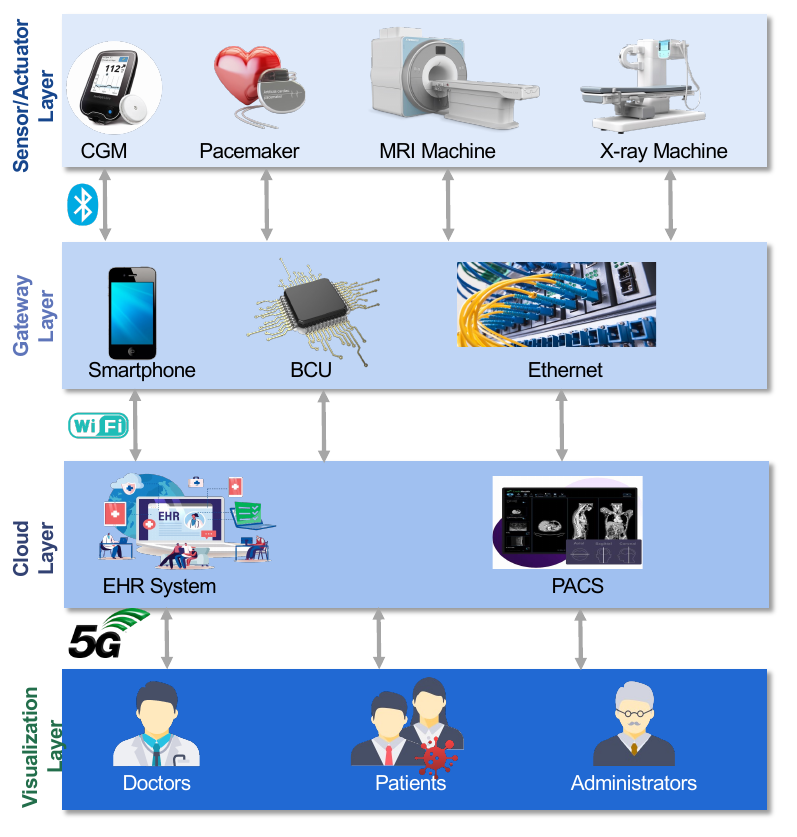}
    \caption{Architecture of IoMT}
    \label{fig:layers}
\end{figure}

Although many legacy medical devices remain unconnected, there is a growing trend of equipping devices with connectivity and embedded intelligence. These modern medical devices interact with physicians, cloud-assisted data centers, hospital management platforms, and data analytics systems, together forming what is termed an Internet of Medical Things (IoMT) system. Most IoMT systems are structured in a layered architecture, typically consisting of four distinct layers (Figure~\ref{fig:layers}, Table~\ref{tab:iomt_summary}). These layers span the entire data lifecycle, from the initial collection of biometric signals to their analysis and visualization by healthcare providers or patients~\cite{jahankhani2019digital,ghubaish2020recent}, ultimately allowing personalized and proactive care.

\begin{table}[H]
\footnotesize
    \centering
    \caption{Summary of IoMT Architecture}
    \label{tab:iomt_summary}
    \renewcommand{\arraystretch}{1.5} 
    \rowcolors{2}{gray!10}{white}
    \begin{tabular}{>{\centering\arraybackslash}p{3cm}>{\centering\arraybackslash}p{3.6cm}>{\centering\arraybackslash}p{3cm}}
        \hline
        \rowcolor{gray!30}
        \textbf{Layer}            & \textbf{Devices/Technologies}                        & \textbf{Functions}                                \\ \hline
        \textbf{Sensor/Actuator Layer}     & Smartwatches, CGMs, MRI machines, Pacemakers        & Collect real-time patient data                    \\ \hline
        \textbf{Gateway Layer}    & Smartphones, Wi-Fi hubs, BCUs              & Aggregate, filter, secure, and transmit      \\ \hline
        \textbf{Cloud Layer}      & Cloud, PACS~\cite{pacs}, EHR, EMR servers                      & Store, analyze, and secure data                   \\ \hline
        \textbf{Visualization Layer} & Dashboards, Mobile Apps, Telemedicine Tools      & Present data to physicians, patients           \\ \hline
    \end{tabular}
    
\end{table}

\begin{itemize}
    \item \textbf{Sensor/Actuator Layer:} This layer includes \textit{first}, wearable sensors (e.g., heart rate monitoring devices, blood glucose monitors); \textit{second}, implantable devices (e.g., cardiac pacemakers, cochlear implants) for monitoring physiological parameters and \textit{third}, on-site medical equipment (e.g., MRI machines, CT scanners, x-ray machines, ventilators, dialysis machines) and \textit{fourth}, actuator devices (e.g., insulin pumps) for drug delivery. Devices may have a purely sensing function, but sometimes contain both sensing and actuation functions. Some degree of interoperability means that a single patient may have distinct sensors and actuators working together to perform a single function.
    \item \textbf{Gateway Layer:} The data collected by the sensors in the sensor/actuator layer are transmitted to the gateway layer using wireless communication protocols (such as Bluetooth Low Energy, ZigBee, Wi-Fi, MedRadio~\cite{medradio}) or wired communication protocols (such as Ethernet). The gateway layer sits between the sensor/actuator layer and the cloud layer. It comprises devices with computational capabilities higher than those of sensors, such as smartphones, body control units (BCUs), or dedicated access points. The gateway layer handles protocol conversion, data filtering, aggregation, and security (such as encryption or decryption) before sending data to the cloud.
    \item \textbf{Cloud Layer:} The gateway layer sends the data (received from the sensor/actuator layer) to the cloud layer using communication technologies such as cellular networks (4G/5G). The cloud layer consists of cloud-based servers (managed by a healthcare provider) and forms the backbone of IoMT, responsible for securely storing and analyzing large volumes of sensor data. It also facilitates access to these data for healthcare providers and patients. The key functions of this layer include: \textit{First}, secure data handling using encryption and authentication to protect transmission of electronic health records (EHRs), also known as Electronic Medical Records (EMRs). \textit{Second}, advanced analytics, including machine learning algorithms, which are applied in the cloud to identify trends, detect health anomalies, and generate actionable insights. Philips HealthSuite~\cite{filips}, for example, is a cloud platform that aggregates and analyzes data from wearable devices to provide actionable insights to patients and physicians. The cloud layer need not be hosted through a third-party cloud provisioning but can be maintained as an on-premises server (private cloud) in the hospital itself.

    \item \textbf{Visualization Layer:} This layer hosts user applications and interfaces. This layer presents the data analyzed (from the cloud layer) to end-users in an understandable and actionable format, facilitating healthcare decision-making. Physicians can access dashboards showing patient health metrics, trends, and alerts, while patients can view simplified reports or notifications on their mobile apps. For example, a telemedicine app might display daily summaries of a patient’s vital signs along with the corresponding physician recommendations.
\end{itemize}


\subsubsection{End-to-End IoMT Workflow: An Illustration in a Hospital Setting}\label{sec:john}

Let us now understand a real-world IoMT system using a  hospital network as a comprehensive example. Consider a patient named John, who is undergoing treatment for diabetes and cardiac problems in a hospital. John’s care involves multiple IoMT devices, systems, and layers. 

\textbf{Sensor/Actuator Layer:} At this layer, various IoMT devices collect real-time health data from John. A Continuous Glucose Monitor (CGM) (e.g., Dexcom G6) monitors John’s blood glucose levels every 5 minutes and transmits data to a smartphone app. The cardiac pacemaker monitors and regulates John’s heart rhythm while sending periodic health data to the hospital gateway. The MRI machine scans John’s heart to assess possible damage caused by diabetes-related complications. An X-ray machine might be used to assess any other concerns, such as respiratory function.

\textbf{Gateway Layer:} The data collected from John by the wearable, implantable, and on-site medical devices are transmitted to their respective gateway layers for intermediate processing. The CGM sends data via Bluetooth to John’s smartphone. The smartphone acts as a hub that aggregates the data and sends them over cellular network to a server provided by the CGM manufacturer. A dedicated BCU acts as the pacemaker's hub, transmitting heart rhythm data securely to the hospital network. MRI and X-ray machines transmit large imaging files via Ethernet to the hospital’s PACS server.

\textbf{Cloud Layer:} John's data can be managed in the hospital's private network or in a remote cloud server. High resolution MRI and X-ray images are stored in the Picture Archiving and Communication System (PACS)~\cite{pacs} server in DICOM (Digital Imaging and Communications in Medicine~\cite{dicom}) format. The EHR/EMR System aggregates patient data (images, data from wearables and implants) into John’s centralized medical record. Data analytics tools in the cloud layer can analyze John's CGM data and pacemaker data to detect episodes of hyperglycemia and arrhythmia. AI-powered tools in the cloud can identify anomalies such as cardiac tissue damage or signs of diabetic cardiomyopathy from the MRI and X-ray images.

\textbf{Visualization Layer:} The analyzed data are presented in a suitable format to the doctor, patient, and hospital administrators. A doctor accesses John’s data through a dashboard integrated with the EHR and the PACS server, which displays heart rate trends, glucose levels, and alerts for irregularities. The PACS server provides high-resolution images (MRI and X-ray) with AI-identified annotations (e.g., tissue damage). The system can potentially detect high-risk events, such as imminent hyperglycemia or arrhythmia, allowing timely adjustments in therapy. John views simplified reports on his mobile app, including daily glucose trends with dietary recommendations, notifications about abnormal heart rhythms, and suggestions to consult with his doctor. Hospital administrators use dashboards to manage on-site devices, track patient status, and ensure resource optimization.

\subsection{How is IoMT Different from a Cyber Physical System?}
A Cyber-Physical System (CPS) is an integrated system that \textit{combines physical components (such as sensors, actuators, and machines) with computational (cyber) components (such as software, algorithms, and networks)} to analyze, optimize, and control processes in the physical world. It relies on a continuous feedback loop in which sensors monitor the physical environment, the computational system processes the data, and actuators adjust the physical system
accordingly. CPS emphasizes the \textit{tight coupling of sensing, computation, and actuation} for precise control.

A hospital’s closed-loop infusion pump system is an example of CPS. The pump’s internal sensors continuously measure blood glucose levels of the patient. Algorithms analyze these glucose readings locally and dynamically adjust the amount of insulin released by the infusion pump (actuator) \textemdash all in real time. After a meal, the patient's blood glucose levels begin to rise. The pump's glucose sensor detects this increase and communicates the data to the insulin pump. The pump increases insulin delivery to prevent hyperglycemia (high blood sugar). Once glucose levels stabilize, the pump gradually reduces the insulin delivery rate to avoid hypoglycemia (low blood sugar). Similarly, if the sensor detects that the patient’s blood glucose is dropping too quickly (a possible precursor to severe hypoglycemia), the insulin pump can automatically stop insulin delivery to prevent further drop in blood glucose levels. In this way, a CPS (here,  closed-loop infusion pump system) uses real-time data to dynamically adapt to changing conditions, ensuring optimal outcomes (in this case, stable blood sugar levels) for users.

IoMT and CPS share similarities, as both integrate physical components with cyber technologies to provide intelligent services. However, they differ in their core objectives, focus, and architectures. Table~\ref{table:summary} highlights how CPS and IoMT play complementary but distinct roles in the advancement of technology and healthcare. 

However,it is important to note that the boundary between CPS and IoMT is not always clearly defined. For example, an insulin pump can be classified as a CPS when it acts as a standalone device with integrated sensors and actuators. When that same device communicates with or is controlled by a remote server, it falls under the broader domain of IoMT. While remote administration of medical devices remains relatively uncommon in clinical settings, it is not unprecedented. The authors have received anecdotal reports of caregivers using unofficial mobile applications to administer insulin remotely, underscoring the evolving nature of the use and classification of medical devices.

\begin{table}[H]
\centering
\caption{Summary of the differences between CPS and IoMT}
\label{table:summary}
\begin{tabular}{lcc}

\hline
\rowcolor{gray!30}
\textbf{Aspect} & \textbf{CPS} & \textbf{IoMT} \\ \hline
\multirow{2}{*}{\textbf{Architecture}} & Tightly coupled     & Connected eco-system of  \\ 
& standalone system & devices and systems\\ \hline

\multirow{2}{*}{\textbf{Primary Objective}} & Real-time control and & Diverse healthcare   \\ 
& monitoring of a specific task  &  management \\ \hline
\textbf{Real-world Example} & Closed-loop infusion pump & Smart healthcare network\\ \hline
\end{tabular}
\end{table}

\section{IoMT Cybersecurity: How is It Different from Traditional IT Security?}

While IoMT builds on the foundational principles of Cyber-Physical Systems (CPS), its interconnected nature, distributed intelligence with a real-time feedback loop, and close link with safety hazards present unique security and privacy issues. Unlike standalone CPS devices, which operate in tightly integrated closed-loop feedback systems, IoMT devices function on diverse, decentralized networks~\cite{fortune}, which include a mix of clinically approved devices (MRI machines) and regular consumer devices (e.g., smartwatch). For example, in a typical hospital in the US, it is reported that more than 3,850 IoMT devices are deployed~\cite{junyper}, which increased significantly due to constraints imposed during the COVID-19 pandemic. This extensive interconnectivity dramatically expands the attack surface, making medical devices, patient data, and critical healthcare infrastructure prime targets for cyber threats.

\begin{figure}[hbt]
    \centering
    \includegraphics[width=0.6\linewidth]{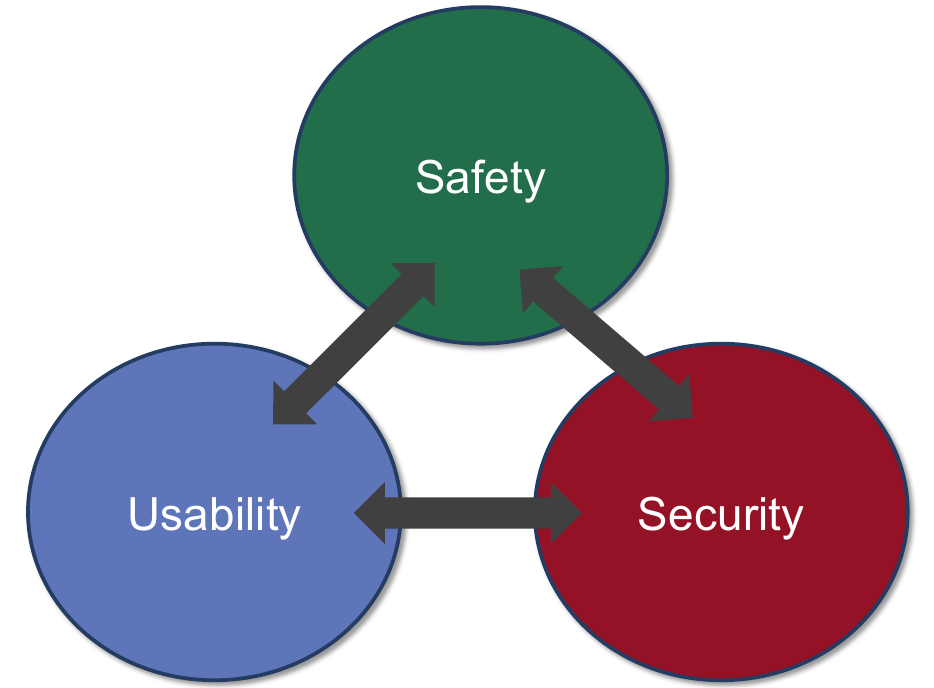}
    \caption{Trade-off Between Various IoMT Objectives}
    \label{fig:usability_safety}
\end{figure}

However, ensuring robust cybersecurity in IoMT is much more challenging than in traditional IT systems. For IoMT, security measures must not only protect sensitive medical data but also ensure that life-saving devices remain accessible, reliable, and functional in critical situations. Furthermore, the interoperability between clinically approved devices (which undergo safety criticality tests) and regular consumer devices (which do not) opens up a wide range of vulnerabilities. More specifically, a security or privacy breach in a consumer device can manifest itself as a safety hazard in a clinical device, both of which remain integrated within an IoMT. The following points highlight these unique challenges, reflecting the intricate interplay between \textit{security}, (patient) \textit{safety}, and \textit{usability} (Fig.~\ref{fig:usability_safety})  in the IoMT ecosystem~\cite{arnabray}.
\begin{enumerate}
    \item Standard security measures such as passwords, biometric scans, or cloud-based authentication, which work in traditional IT settings, can be impractical or even dangerous in medical settings.  For example:
    \begin{exmp}
        In the emergency departments in hospitals, ventilators are crucial for life support. Requiring passwords on such devices would enhance cybersecurity, but waste precious seconds or endanger the patient's life in emergency scenarios, especially if the caregiver forgets the password.
    \end{exmp}
    \begin{exmp}
        Biometric authentication methods such as fingerprint or facial recognition also do not work well in a medical setting due to the use of hand gloves and surgical masks.
    \end{exmp}
        \begin{exmp}
            Patients often suffer from conditions such as poor vision and arthritic fingers that can make it difficult for them to enter long alphanumeric passwords before using any medical app or device.
        \end{exmp}  
        \begin{exmp}
            Healthcare authorities refrain from adopting security decisions that cause changes in the workflow for users \textemdash clinicians, patients, or service technicians \textemdash because they fear that users may struggle with the new workflow.
        \end{exmp}
    
    Therefore, cybersecurity experts must understand the interplay of these clinical nuances while designing IoMT cybersecurity.  

    \item Threat modeling is essential to identify how cybersecurity risks can compromise patient safety and privacy. Modeling risks to patient safety from cybersecurity threats is fundamentally different from assessing risks associated with mechanical, electrical, software, or human factors failures. We cite the following examples to better explain these differences:
                    \begin{exmp}
                        Mechanical failure of a ventilator, such as a seizing motor, directly stops airflow to a patient, causing immediate and observable harm. Similarly, an electrical failure can cause the device to shut down due to a power surge, which is often detectable and can be mitigated with backup systems, such as uninterruptible power supplies.
                    \end{exmp}

                    \begin{exmp}
                        A ransomware attack targeting the same ventilator may not cause an immediate shutdown but could encrypt its control system, leaving the hospital unable to adjust settings or ensure proper operation. Unlike mechanical failures, cybersecurity threats can result in delayed harm, as they often prevent timely intervention or mask the issue entirely.
                    \end{exmp}
                    \begin{exmp}
                        An MRI machine could fail to initialize due to a bug in its operating system, preventing scans from being performed. This type of failure is typically logged and can be addressed by rebooting or patching the system. However, a malicious actor exploiting a vulnerability in the network could alter the diagnostic images produced by the MRI machine, leading to incorrect interpretations of patient conditions. This type of attack is not only harder to detect, but also has broader implications, as it undermines the trust in diagnostic accuracy.
                    \end{exmp}
                    \begin{exmp}
                        A fitness tracker might stop measuring a patient’s heart rate due to a sensor malfunction, alerting the user with a clear error notification.
                    \end{exmp}
                    \begin{exmp}
                        A cybercriminal who intercepts unencrypted Bluetooth data from the same tracker could manipulate the readings, making it appear that the patient's heart rate is dangerously high or low, potentially leading to unnecessary or even harmful medical interventions.
                    \end{exmp}
    
    Many key points set threat modeling for IoMT cybersecurity apart from traditional threat modeling. Traditional risks (mechanical, electrical, or software) are often predictable and detectable, with standardized mitigation strategies. Cybersecurity risks are more dynamic and may involve hidden attacks (e.g., ransomware or data tampering) that exploit interconnected systems. Mechanical or electrical failure is usually isolated to a specific device, whereas a cybersecurity breach can spread across interconnected devices, affecting multiple systems simultaneously. Cybersecurity threats often require collaboration between medical device manufacturers, hospitals, and IT teams, while traditional failures are typically resolved by engineers or maintenance staff.

    \item Medical devices often remain in use for decades, even after manufacturers stop supporting them, primarily because of the devices' high cost. Hospitals often choose to maintain these expensive devices, such as CT scanners and MRI machines, by relying on third-party service providers rather than replacing them. However, these service providers cannot provide software updates such as security patches, leaving these devices vulnerable to cybersecurity risks.

    \item Medical device designers have traditionally designed devices for non-networked, offline use. Later, when these legacy devices are integrated into the connected digital health ecosystem (say, by adding a wireless card to a medical device), they open up control pathways that were previously unavailable, making them vulnerable to data breaches. Older medical devices lack the computational capacity to perform cryptographic operations without sacrificing functionality and may not have the memory needed for encryption software. Hardware from previous generations also lacks the ability to store secure cryptographic keys. As a result, such medical devices are often found to transmit patient data without encryption. 
    
    The recommended approach to secure such legacy medical devices is hardware isolation, with separate chips managing communication security and clinical functions, reducing the risk of connectivity-based attacks. Unlike the smartphone industry, where frequent hardware updates are common, the medical device industry faces higher regulatory hurdles for software verification, making it costly and time-consuming to upgrade hardware solely for security. 

    \item Due to the high costs of medical devices, healthcare organizations, such as hospitals, are generally reluctant to pay extra for devices that offer only additional security hardware, especially when they do not provide new therapeutic or diagnostic benefits. So, in the medical device industry, a hardware upgrade solely for security purposes is often not considered commercially viable, unlike the consumer market, where frequent upgrades are driven by customer demand for better performance and features.

    \item Due to the long service time of medical devices, manufacturers often need to ensure that their new devices remain compatible with their older, less secure models that are still in use. This presents a unique challenge. To understand this, let us consider a common scenario in which a medical device manufacturer is set to release a new line of implantable devices and bedside monitoring units. For this new generation, the manufacturer has added security to its Bluetooth communication so that the implantable device and the bedside monitors communicate over an encrypted and authenticated channel. However, many hospitals still rely on older bedside monitors, which cannot communicate securely with the latest heart monitors due to the lack of encryption support (software as well as hardware) in older models. The manufacturer has two available choices: either replace all the outdated, still-in-use bedside monitors with newer, secure models or design the new implantable devices such that they can also connect with the old bedside units through an unsecured channel. While the first choice would incur huge cost, the second option would weaken the security of the entire product line. Thus, enforcing a strict security standard for new medical devices can very likely lead to compatibility issues, posing both clinical and business risks.

    \item The application of security patches to medical devices in healthcare networks involves numerous complications.
    Most hospital administrators do not allow direct internet connections to their devices, through which security patches can be obtained remotely. Many medical devices, especially legacy ones, in use do not have the cryptographic hardware and/or software to verify the authenticity and integrity of the security patches. Unlike consumer devices, medical devices have the risk of interrupting therapy and affecting patient safety when receiving security patches. For example, a vulnerable patient may have to visit the clinic to get a security patch applied on his or her implanted device. The doctor can advise whether this would be feasible depending on the patient's condition. So, it is the healthcare providers and patients who ultimately decide whether or when to apply patches. Medical device manufacturers cannot simply enforce patch updates, but can encourage adoption through education and outreach. Applying patches on medical devices is more complex and time-consuming than in traditional software, as errors can directly impact patient safety. If a medical device manufacturer applies an incorrect patch, it could compromise device functionality, possibly causing harm to patients. To avoid risks, manufacturers rigorously test third-party patches, but this delays the update process. This delay benefits attackers, who may exploit the time between when a vulnerability is publicly announced and when the patch is fully deployed.

    \item Hospitals purchase and deploy medical devices on their network and are often held responsible for their security by patients. However, hospitals typically lack information about the cybersecurity posture or vulnerabilities of these devices. Medical device manufacturers, on the other hand, focus on manufacturing and selling devices and are minimally involved in post-deployment security. This division of responsibilities often creates gaps in cybersecurity, as neither party has full visibility or control over all aspects of security. For example, if a hospital’s radiology machine is compromised due to unauthorized remote access by a former staff member, the manufacturer could attribute this to poor access controls in the hospital. In contrast, the hospital might claim that the manufacturer should have designed the machine to limit remote access. In essence, ensuring clear and effective cybersecurity in medical settings remains a complex challenge due to the differing priorities and capabilities between medical device manufacturers and hospitals.
\end{enumerate}

\section{State of IoMT Cybersecurity: Incidents, Trends, and Impact}


This section presents a chronological overview of real-world events that have shaped the landscape of IoMT cybersecurity. By examining these historical incidents, we gain valuable insights into the evolving vulnerabilities of IoMT systems. These events have had significant repercussions on the adoption of IoMT, prompting regulatory authorities to incorporate cybersecurity assessments into their approval processes~\cite{Singapore_MD_Cyber, US_FDA_Cyber, mdcg_europe_cybersecurity}. In fact, it is no longer uncommon for medical devices to be recalled due to cybersecurity concerns~\cite{insulin_security_recall, abbott}.

\subsection{Timeline of Major Incidents in IoMT Cybersecurity}
\begin{itemize}
    \item \textbf{2007}: Former US Vice President Dick Cheney revealed that, when his cardiac pacemaker was replaced in 2007, his cardiologist had disabled the device's wireless function as there was intelligence that terrorists could use it to hack the device and send fatal shocks to his heart~\cite{dick}.
    \item \textbf{2008}: Halperin \textit{et al.} demonstrated that implantable cardiac defibrillators could be hacked using software-defined radios~\cite{halperin2008security}.
    \item \textbf{2010}: In a hearing on reviewing information security at the US Department of Veterans Affairs, the Honorable Roger W. Baker said, ``Over 122 medical devices have been compromised by malware over the last 14 months. These injections have the potential to greatly affect the world-class patient care that is expected by our customers''~\cite{veterans}.
    \item \textbf{2011}: Jay Radcliffe, a security researcher, demonstrated cybersecurity vulnerabilities in his insulin pump at the Black Hat conference~\cite{black}.
    \item \textbf{2012}: Barnaby Jack of security vendor IOActive found that pacemakers from various manufacturers could be remotely manipulated to deliver a lethal 830-volt shock using a laptop within 50 feet — a vulnerability stemming from flawed software programming by medical device companies~\cite{830}.
    \item \textbf{2014}: The US FDA issued pre-market guidelines for cybersecurity~\cite{fda14}.
    \item \textbf{2015}: Anthem Inc., the then second-largest health insurer in the US,  was hit by a major cyberattack in which 78.8 million personal health records were stolen~\cite{ant}. It remains one of the largest data breaches to date.
    \item \textbf{2015}: The U.S. Food and Drug Administration (FDA) advised hospitals to stop using Hospira Inc's Symbiq infusion system due to a security vulnerability that could allow cyberattackers to remotely control the device. This advisory came about 10 days after the U.S. Department of Homeland Security (DHS) issued a warning about the same vulnerability. This marked the first instance in which the FDA recommended discontinuing the use of a medical device due to a cybersecurity risk~\cite{symbiq}.
    \item \textbf{2015}: The US Congress passed the Cybersecurity Act. Section 405 of the Act laid out steps for strengthening the cybersecurity of the healthcare industry, including the establishment of the Health Care Industry Cybersecurity (HCIC) Task Force~\cite{hcic}. The HCIC, in its report, criticized the medical device manufacturers for ignoring cybersecurity.
    \item \textbf{2016}: Hollywood Presbyterian Medical Center in Los Angeles was attacked with ransomware. The hospital eventually paid the hackers \$17,000 in Bitcoins to regain access~\cite{hollywood}.
    \item \textbf{2016}: In a letter to the chief executives of Johnson \& Johnson, GE Healthcare, Siemens, Medtronic and Philips, which collectively controlled more than a quarter of the global medical device market, then-Senator Barbara Boxer expressed concern over device cybersecurity and urged them to share their plans to deal with it~\cite{boxer}. Independent security researchers discovered that a specific infusion system had vulnerabilities allowing unauthorized users to access the device via a hospital’s network, potentially enabling them to control the device, alter dosage levels, and endanger patient safety.
    \item \textbf{2016}: St. Jude Medical's stock price fell sharply after reports that its implantable heart devices were susceptible to cyberattacks emerged in the public~\cite{jude}. In the same year, US FDA issued post-market cybersecurity guidance~\cite{fda16}.
    \item \textbf{2017}: About 500,000 pacemakers, all made by the medical device company Abbott and sold under the St Jude Medical brand, were recalled by the US FDA for a critical firmware update to patch security flaws~\cite{abbott}. 
    \item \textbf{2017}: The National Health Service in the United Kingdom and numerous other healthcare providers around the globe were adversely impacted by the WannaCry ransomware attack~\cite{cry}. The impacts included critical medical machinery, like MRI scanners, rendered unusable by the medical staff, doctors unable to administer medication as they were blocked from accessing the patients' medical records, and the  emergency units closed, to name a few. The WannaCry attack exploited a vulnerability in the SMB file-sharing protocol. Vulnerability to such ransomware attack is amplified by the presence of connected devices and open ports.
    \item \textbf{2018}: In Black Hat, researcher Billy Rios revealed multiple life-threatening vulnerabilities in Medtronic's software delivery network used for updating its pacemaker programmers~\cite{medtronic}.
    \item \textbf{2018}: Philips identified nine cybersecurity vulnerabilities in its e-Alert MRI monitoring system. According to CISA, these vulnerabilities could allow attackers to input unexpected commands, execute arbitrary code, display unit information, or potentially cause the e-Alert system to crash~\cite{mri}.
    \item \textbf{2018}: The US FDA issued the final version of pre-market cybersecurity guidance, this time making its regulatory expectations more stringent~\cite{fda18}.
    \item \textbf{2019}: Two significant vulnerabilities were discovered in Medtronic's Conexus telemetry protocol, affecting MyCarelink monitors, CareLink programmers, and several implanted cardiac devices. The critical vulnerability (CVE-2019-6538), rated at 9.3 on the CVSS scale, allowed attackers with close-range access to intercept and modify device communications due to a lack of authentication controls. A second vulnerability (CVE-2019-6540), rated medium severity, involved clear-text transmission of sensitive data, making it vulnerable to interception~\cite{hipaa}.
    \item \textbf{2020}: The US Department of Homeland Security issued a cybersecurity advisory on the MyCareLink product line of Medtronic. Later, Medtronic released a firmware update to address the issues~\cite{mycare}.
    \item \textbf{2021}: The Owlet Smart Sock \textemdash a wearable heart monitor for infants \textemdash was removed from the market after the US FDA issued a warning letter citing regulatory violations~\cite{owl}.
    \item \textbf{2023}: Medical device manufacturer BD issued a bulletin disclosing a password vulnerability in one of its infusion pumps, which could potentially allow access to personal information~\cite{bd}.
\end{itemize}

\subsection{Key Statistics and Insights}
To supplement the timeline above, we present key statistics that highlight the scale and severity of cybersecurity threats to IoMT systems. These statistics provide a clear, data-driven view of how cyberattacks affect the Internet of Medical Things (IoMT), highlighting their financial, operational, and privacy-related consequences for healthcare providers, medical device manufacturers, and patients.

\begin{figure}[H]
    \centering
    \includegraphics[width=0.9\textwidth]{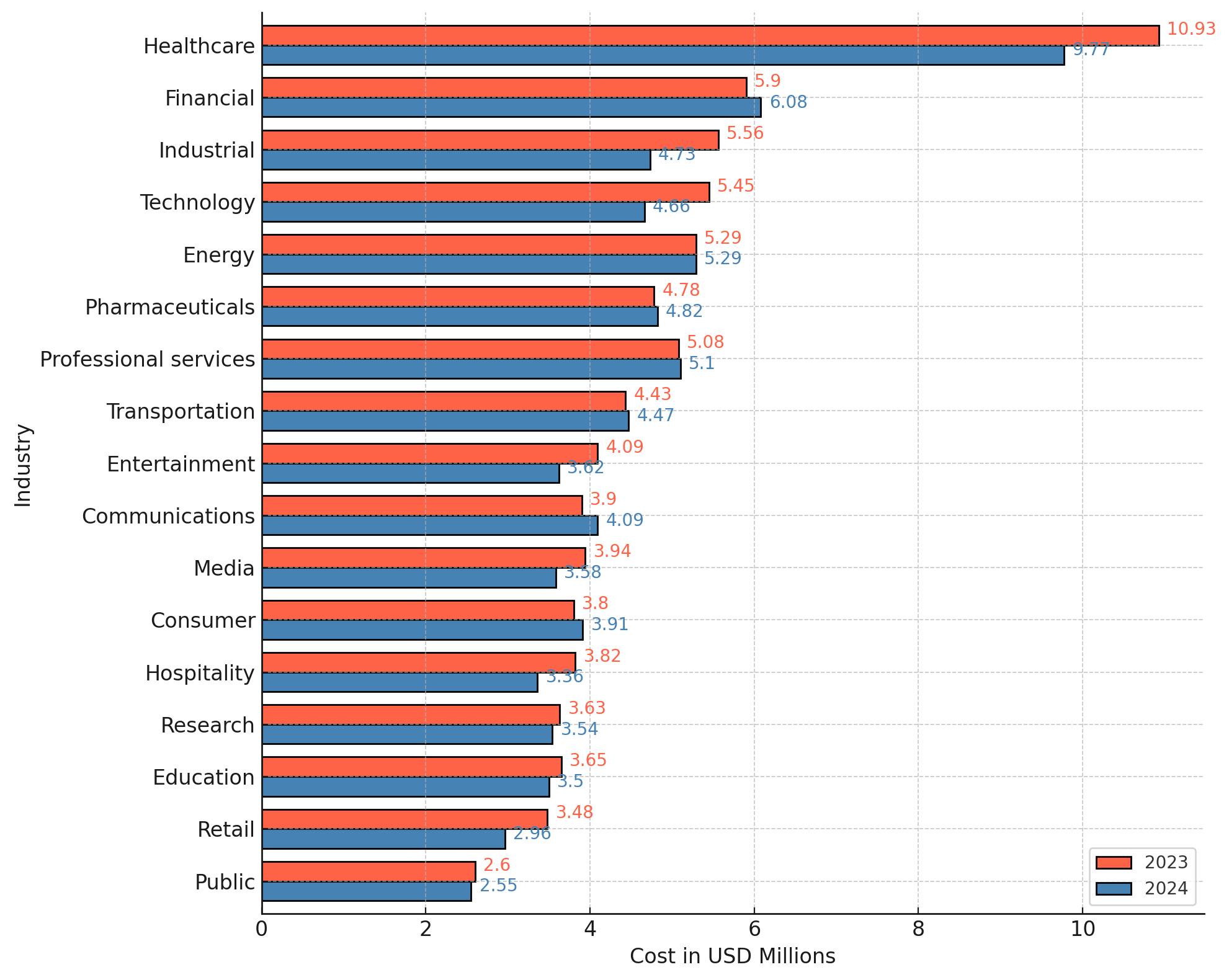}
    \caption{Cost of Data Breach by Industry (2023 Vs 2024)~\cite{ibmsec}.}
    \label{ibmpic}
\end{figure}

According to the Department of Health and Human Services, US Office of Information Security, approximately 385 million patient records were potentially exposed to data breaches between 2010 and 2022~\cite{hhs}. While this results from a typical IT security breach, it has serious implications in IoMT cybersecurity practices. In the black market, selling the patient's health record fetches significantly more money than selling financial information. While the data of a stolen credit card is sold for a few cents on the black market~\cite{dmag}, medical record of a patient is estimated to be \$250 as per one study~\cite{trustwave} and ranging from \$1 to \$1000 (depending on how complete the record is) according to the studies~\cite{experian,forbes,becker}. According to IBM Security, healthcare continues to be the industry with the most costly data breaches in 2024, for the $13^{th}$ consecutive year. Figure~\ref{ibmpic} shows that the average total cost of a data breach in the healthcare industry is USD 9.77 million in 2024~\cite{ibmsec}. 

A report by Statista estimates that the number of hospitals in the world will reach 166,548 by 2029. The average number of connected medical devices per hospital bed, according to the HIPAA (Health Insurance Portability and Accountability Act) Journal, is approximately 10 to 15. This puts the number of connected medical devices in the world at 1.67 million by 2029~\cite{secintel}. The medical data security firm CloudWave (formerly Sensato) found an average of 6.2 vulnerabilities per medical device. To make the situation even worse, 60\% of these devices tend to be at the end of their life cycles, with no patches or upgrades available~\cite{sensato}. According to data from the CyberPeace Institute, a cyberattack on a healthcare system results in an average of 19 days of interrupted patient care~\cite{cyberpeace}. A 2018 study examining nearly 5,000 medical devices with software components found that only 2.13\% of their manuals included cybersecurity information~\cite{stern2019cybersecurity}. A 2023 study published in Nature Scientific Reports found that medical devices remain exposed to cybersecurity vulnerabilities for an average of 3.2 years even if they receive the security patch the day the vulnerabilities are discovered~\cite{bracciale2023cybersecurity}. It is evident that the adoption of IoMT is occuring much faster than the corresponding safety analysis, thus leading to a much more vulnerable IoMT fabric than now.

\section{Cybersecurity Attacks in IoMT: A Survey and Taxonomy}
The cybersecurity incidents and statistics presented in the previous section highlight the growing sophistication of cyber threats in the IoMT domain. Cyberattacks on IoMT systems can range from targeted attacks on life-critical medical devices to large-scale breaches that expose millions of sensitive patient records. To systematically understand these security challenges, we conducted an extensive review of the existing research literature, analyzing state-of-the-art attacks on networked medical devices. 

\subsection{Cyberattack Taxonomy: State-of-the-Art}

\begin{figure}[H]
    \centering
    \includegraphics[width=\linewidth]{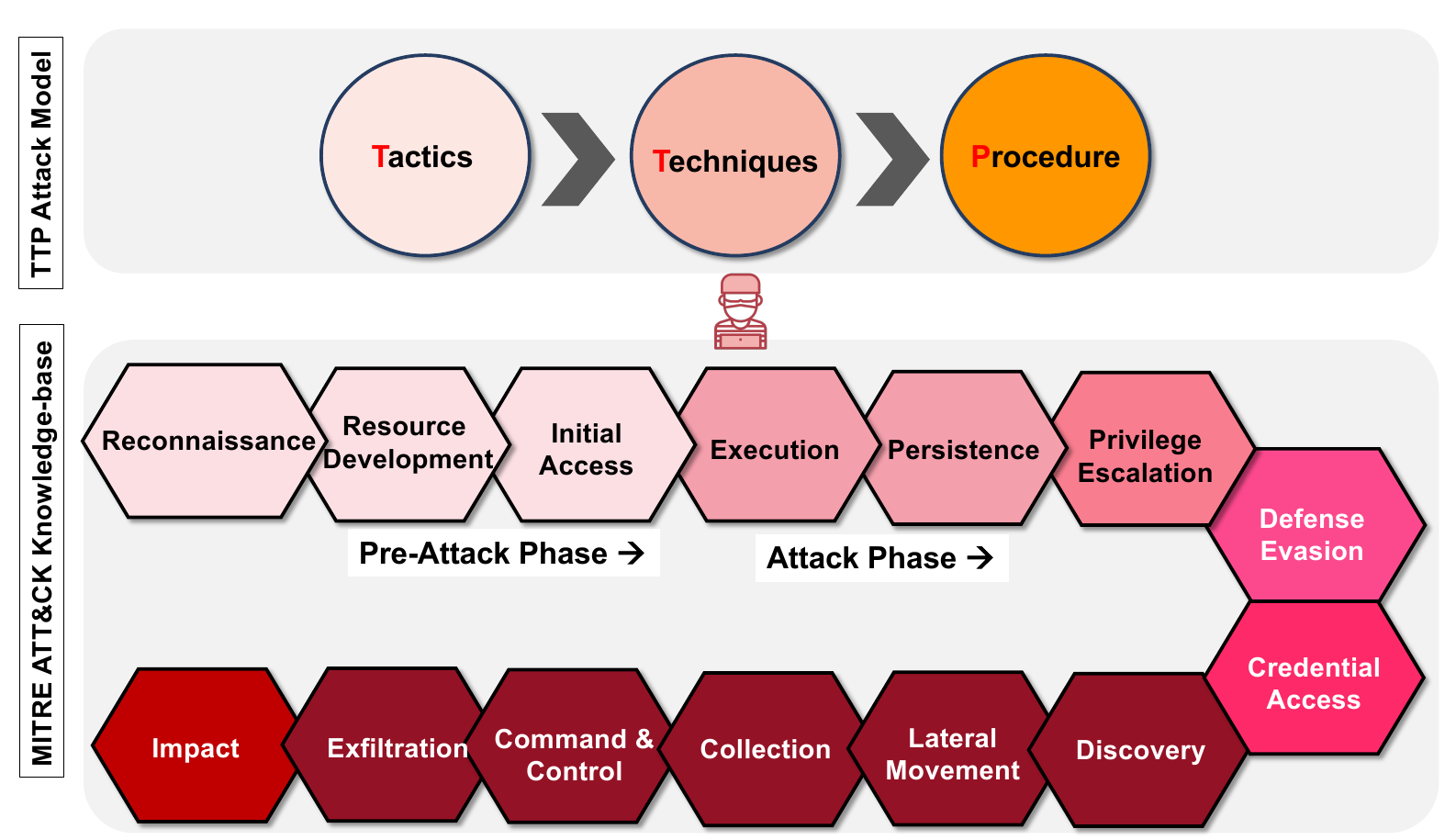}
    \caption{Cybersecurity Attack Models}
    \label{fig:attack_framework}
\end{figure}

Several prominent cyberattack taxonomies are currently in use, each offering a different level of granularity and focus. These include the Threat-Vulnerability-Risk (TVR) model and the Tactics, Techniques, and Procedures (TTPs) framework. Such taxonomies often build upon more detailed threat modeling approaches, such as the STRIDE model — Spoofing, Tampering, Repudiation, Information Disclosure, Denial of Service, and Elevation of Privilege — which is adaptable across various application domains. A more comprehensive and widely adopted taxonomy is the MITRE ATT\&CK framework~\cite{mitre_attack}, which categorizes attacker behavior in a structured and systematic way. These frameworks, illustrated in Fig.~\ref{fig:attack_framework}, map the complete lifecycle of an attack — commonly referred to as the Cyber Kill Chain~\cite{cyber-kill-chain} — and are particularly valuable for security practitioners engaged in threat detection, mitigation, and incident response.

Several domain-specific cyberattack frameworks have also been developed to address the unique needs of different sectors. In the automotive domain, the TVR model has evolved into the Threat Analysis and Risk Assessment (TARA) framework, which aligns closely with the ISO/SAE 21434 automotive cybersecurity standard. For space and aviation security, the TTP methodology has been extended into the Space Attack Research and Tactic Analysis (SPARTA) framework. In SPARTA, known vulnerabilities and exploits are mapped to defined attack states, and a security score is assigned accordingly. SPARTA identifies nine distinct attack states: Reconnaissance, Resource Development, Initial Access, Execution, Persistence, Defense Evasion, Lateral Movement, Exfiltration, and Impact. This structure closely mirrors the MITRE ATT\&CK framework, which defines 14 stages in the attack lifecycle. The MITRE framework has also been recently extended to cover AI-specific threats under the ATLAS initiative~\cite{mitre_atlas}. Since these domains significantly overlap with traditional IT infrastructures, risk scoring is often aligned with established standards like ISO/IEC 27001 (information security management) and NIST SP 800-53 (security and privacy controls for information systems). In the following section, we present an extended TTP-based taxonomy specifically tailored for IoMT.


\subsection{Proposed Cyberattack Taxonomy for IoMT}
In our study, we found that these attacks exploited a wide array of attack surfaces and vulnerabilities, each at different layers of the IoMT architecture (Figure~\ref{fig:layers}). As shown in Figure~\ref{fig:layers}, the IoMT architecture is made up of four primary layers: sensor/actuator layer, gateway layer, cloud layer, and visualization layer. Each layer introduces unique security challenges due to the heterogeneous nature of medical devices, communication protocols, and data processing mechanisms that operate within them. Based on these findings, we propose a structured taxonomy that classifies the cyberattacks 
 and their associated attack surfaces and vulnerabilities according to the specific layers (within the IoMT architecture) in which they reside.

Note that our taxonomy treats communication channels between consecutive IoMT layers as distinct attack surfaces. This refined approach improves the granularity and precision of our classification by identifying vulnerabilities not only within individual layers, but also at the critical interfaces where data are transmitted between them.

\subsection{Sensor/Actuator Layer: Attack Surfaces and Vulnerabilities}

\begin{table}[H]
    \centering
    \caption{Taxonomy of Attacks in IoMT: Sensor/Actuator Layer}
    \label{tab:tax_sensor_lyr}
    \renewcommand{\arraystretch}{1} 
    \rowcolors{2}{gray!10}{white}
    \begin{adjustbox}{width=1.2\textwidth,center=\textwidth}
    \begin{tabular}{>{\centering\arraybackslash}p{2.3cm}>{\centering\arraybackslash}p{2.5cm}>
    {\centering\arraybackslash}p{3.2cm}>
    {\centering\arraybackslash}p{1.7cm}>
    {\centering\arraybackslash}p{2cm}>
    {\centering\arraybackslash}p{2.7cm}>
    {\centering\arraybackslash}p{3cm}}
        \hline
        \rowcolor{gray!30}
        \textbf{Attack Surface} 
        & 
        \textbf{Vulnerability} ($V_S$XX) 
        & 
        \textbf{Attack Methodology} 
        & 
        \textbf{Attacker's Access} 
        & 
        \textbf{Attack Stage} 
        & 
        \textbf{Target} 
        & 
        \textbf{Attack Impact} 
        \\
        \hline

        \textbf{Sensing, Pacing leads}
        &
        \textcolor{magenta}{\textbf{Susceptible to electro-magnetic interference}} ($V_S01$)
        &
        Introduce differential voltages in the leads~\cite{2019survey_ref46}
        &
        Remote, up to 1.5 meters
        &
        Post-deployment phase
        &
        Implantable pacemaker, defibrillator (e.g., Medtronic Adapta)
        &
        Pacemaker prevented from delivering pacing signals, induced false readings
        \\
        \hline

        \textbf{USB ports, CD/DVD drives}
        &
        \textcolor{magenta}{\textbf{Unrestricted access to USB port, CD/DVD drive}} ($V_S02$)
        &
        Network mapping and vulnerability assessments through Nmap and OpenVAS~\cite{2019survey_ref90}
        &
        Physical access to the ports
        &
        Post-deployment phase
        &
        On-site networked medical equipment
        &
        Privacy breach, malware introduction
        \\
        \hline
        \textbf{Operator environment}
        &
        \textcolor{magenta}{\textbf{Malicious app in smartwatch worn by the operator}} ($V_S03$)
        &
        Install a malicious app on the smartwatch~\cite{2019survey_ref81}
        &
        Indirect access to the medical device via smartwatch
        &
        During the regular use of the medical device
        &
        Medical device (with keypad input) that requires PIN for access or configuration
        &
        Attacker can get administrator access
        \\
        \hline

        \multirow{2}{*}{\textbf{Software}}
        &
        \textcolor{magenta}{\textbf{Unpatched software}} ($V_S04$)
        &
        Network mapping and vulnerability assessments  through Nmap and OpenVAS~\cite{2019survey_ref90}
        &
        Ability to exploit known vulnerabilities
        &
        Post-deployment phase
        &
        On-site networked medical equipment
        &
        Gaining unauthorized access, stealing patient data, injecting malware
        \\
        \cline{2-7}
        &
        \textcolor{magenta}{\textbf{Out-of-bounds write (CWE-787)}}  ($V_S05$)
        &
        No attack shown in CVE-2021-27410~\cite{CVE-2021-27410}
        &
        Not mentioned
        &
        Post-deployment phase
        &
        Welch Allyn Connex Vital Signs Monitor (CVSM)
        &
        Corruption of data or code execution
        \\
        \cline{2-7}
        &
        \textcolor{magenta}{\textbf{Buffer overflow}}  ($V_S06$)
        &
        No attack shown in  CVE-2017-12718~\cite{CVE-2017-12718}
        &
        Remote access
        &
        Post-deployment phase
        &
        Smiths Medical Wireless Syringe Infusion Pump
        &
        Remote code execution on the target device
        \\
        \hline
        \textbf{User access}
        &
        \textcolor{magenta}{\textbf{Hardcoded passwords}} ($V_S07$)
        &
        No attack shown in CVE-2020-12039~\cite{CVE-2020-12039}
        &
        Physical access
        &
        Post-deployment phase
        &
        Baxter Sigma Spectrum Infusion Pumps
        &
        Unauthorized access to device settings, patient data, network configuration
        \\
        \hline
        
        \textbf{Configuration file(s)}	
        &
        \textcolor{magenta}{\textbf{Passwords stored in configuration file}} ($V_S08$)
        &
        No attack shown in CVE-2017-12723~\cite{CVE-2017-12723}
        &
        Remote network access without prior authentication
        &
        Post-deployment phase
        &
        Smiths Medical Wireless Syringe Infusion Pump
        &
        Unauthorized access
        \\
        \hline      

        \multirow{2}{*}{\makecell{\textbf{Operating}\\\textbf{System}}}
        &
        \textcolor{magenta}{\textbf{Privilege escalation (CWE-269)}} ($V_S09$)
        &
        No attack shown in  CVE-2021-32025~\cite{CVE-2021-32025}
        &
        Not mentioned
        &
        Post-deployment phase
        &
        QNX Neutrino Kernel in QNX
        &
        Unauthorized access
        \\
        \cline{2-7}
        &
        \textcolor{magenta}{\textbf{Outdated operating system (e.g., Windows XP)}} ($V_S10$)
        &
        Network mapping and vulnerability assessments performed through Nmap and OpenVAS~\cite{2019survey_ref90}
        &
        Exploit known vulnerabilities
        &
        Post-deployment phase
        &
        On-site networked medical equipment
        &
        Gaining unauthorized access, stealing patient data, injecting malware
        \\
        \hline
                
    \end{tabular}
    \end{adjustbox}
    
\end{table}

\setcounter{table}{3}

\setcounter{table}{3}

\begin{table}[H]
    \centering
    \caption{(Continued) Taxonomy of Attacks in IoMT: Sensor/Actuator Layer}
    \label{tab:tax_sensor_lyr}
    \renewcommand{\arraystretch}{1.5} 
    \rowcolors{2}{gray!10}{white}
    \begin{adjustbox}{width=1.2\textwidth,center=\textwidth}
    \begin{tabular}{>{\centering\arraybackslash}p{2.3cm}>{\centering\arraybackslash}p{2.5cm}>
    {\centering\arraybackslash}p{3cm}>
    {\centering\arraybackslash}p{1.7cm}>
    {\centering\arraybackslash}p{2cm}>
    {\centering\arraybackslash}p{2cm}>
    {\centering\arraybackslash}p{2.7cm}}
        \hline
        \rowcolor{gray!30}
        \textbf{Attack Surface} 
        & 
        \textbf{Vulnerability} ($V_S$XX)
        & 
        \textbf{Attack Methodology} 
        & 
        \textbf{Attacker's Access} 
        & 
        \textbf{Attack Stage} 
        & 
        \textbf{Target} 
        & 
        \textbf{Attack Impact} 
        \\
        \hline
        \textbf{Network ports}
        &
        \textcolor{magenta}{\textbf{Open ports}} ($V_S11$); \textcolor{magenta}{\textbf{Default or weak passwords}} ($V_S12$)
        &
        Network mapping and vulnerability assessments performed through Nmap and OpenVAS~\cite{2019survey_ref90}
        &
        Remote attacker via open network ports
        &
        Post-deployment phase
        &
        On-site networked medical equipment
        &
        Unauthorized remote access, malware installation
        \\
        \hline
        \rowcolor{gray!10}
        \textbf{Antivirus}
        &
        \textcolor{magenta}{\textbf{Missing antivirus protection or outdated virus signatures}} ($V_S13$)
        &
        Network mapping and vulnerability assessments performed through Nmap and OpenVAS~\cite{2019survey_ref90}
        &
        Attacker can inject malware in the medical device
        &
        Post-deployment phase
        &
        On-site networked medical equipment
        &
        DoS, ransomware
    \\
    \hline
        
        \multirow{3}{*}{\textbf{Firmware}}
        &
        \textcolor{magenta}{\textbf{Read, write accesses to firmware memory}} ($V_S14$); \textcolor{magenta}{\textbf{missing signature verification}} ($V_S15$)
        &
        Installing custom firmware by boot bypass~\cite{2019survey_ref19, 2019survey_ref77}
        &
        Physical access to BOOT0 pin and USB port
        &
        Post-deployment, maintenance phase
        &
        Nike\texttt{+} Fuelband (Wearable Device)
        &
        Malicious firmware injection, unauthorized device control
        \\
        \cline{2-7}
        &
        \textcolor{magenta}{\textbf{Firmware integrity check through CRC}} ($V_S16$)
        &
        Falsify CRC table to bypass verification~\cite{2019survey_refAED45}
        &
        Physical or network access to the device
        &
        Post-deployment phase
        &
        Automated external defibrillator (AED) 
        &
        Unauthorized firmware modifications
        \\
        \hline             
        \multirow{4}{*}{\makecell{\textbf{Network}\\\textbf{interface}}}
        &
        \textcolor{magenta}{\textbf{Factory account with a hardcoded password}} ($V_S17$)
        &
        No public exploitation of this security vulnerability is known~\cite{CVE-2018-4846}
        &
        Successful exploitation requires no user interaction
        &
        Post-deployment phase
        &
        Siemens RAPIDLab and RAPIDPoint blood gas analyzers
        &
        Unauthorized remote access over port 8900/TCP
        \\
        \cline{2-7}

        &
        \textcolor{magenta}{\textbf{Improper Access Control: No authentication for FTP connections}} ($V_S18$)
        &
        No attack shown in  CVE-2017-12720~\cite{CVE-2017-12720}
        &
        Not mentioned
        &
        Post-deployment phase
        &
        Smiths Medical Wireless Syringe Infusion Pump
        &
        Unauthorized remote access to the device, potentially compromising patient safety
        \\
        
        \cline{2-7}
        &
        \textcolor{magenta}{\textbf{Improper Certificate Validation}} ($V_S19$)
        &
        No attack shown in  CVE-2017-12721~\cite{CVE-2017-12721}
        &
        Remote access
        &
        Post-deployment phase
        &
        Smiths Medical Wireless Syringe Infusion Pump
        &
        Man-in-the-Middle (MitM) attack
        \\
        \hline        
    \end{tabular}
    \end{adjustbox}
    
\end{table}

\subsection{Between Sensor/Actuator Layer and Gateway Layer: Attack Surfaces and Vulnerabilities}

\begin{table}[H]
    \centering
    \caption{Taxonomy of Attacks in IoMT: Between Sensor and Gateway Layer}
    \label{tab:tax_sensor2gateway_lyr}
    \renewcommand{\arraystretch}{1.5} 
    \rowcolors{2}{gray!10}{white}
    \begin{adjustbox}{width=1.2\textwidth,center=\textwidth}
    \begin{tabular}{>{\centering\arraybackslash}p{2.7cm}>{\centering\arraybackslash}p{3cm}>
    {\centering\arraybackslash}p{3cm}>
    {\centering\arraybackslash}p{1.7cm}>
    {\centering\arraybackslash}p{2cm}>
    {\centering\arraybackslash}p{2cm}>
    {\centering\arraybackslash}p{2.8cm}}
        \hline
        \rowcolor{gray!30}
        \textbf{Attack Surface} 
        & 
        \textbf{Vulnerability} ($V_{SG}$XX)
        & 
        \textbf{Attack Methodology}
        & 
        \textbf{Attacker's Access} 
        & 
        \textbf{Attack Stage} 
        & 
        \textbf{Target} 
        & 
        \textbf{Attack Impact} 
        \\
        \hline
        \rowcolor{gray!10}
        
        \textbf{DICOM communication from imaging devices}
        &
        \textcolor{magenta}{\textbf{The DICOM standard supports encryption but is not enforced}} ($V_{SG}01$)
        &
        Network mapping and vulnerability assessments performed using Nmap and OpenVAS ~\cite{2019survey_ref90}
        &
        Attacker can intercept  unencrypted DICOM images
        &
        Post-deployment phase
        &
        On-site imaging equipment
        &
        Unauthorized access to patient scans
        \\
        \hline         
        \textbf{Communication between medical mannequin and its controller (laptop)}
        &
        \textcolor{magenta}{\textbf{Wi-Fi Protected Setup PIN can be discovered by a brute-force attack}} ($V_{SG}02$); \textcolor{magenta}{\textbf{Outdated Wi-Fi 802.11 standard}} ($V_{SG}03$)
        &
        Used BackTrack 5 to scan nearby wireless networks and identified the mannequin’s access point using its MAC address and channel~\cite{2019survey_ref63} DoS through repeated de-authentication
        &
        Local (proximity) wireless access
        &
        Post-deployment phase
        &
        iStan medical mannequin
        &
        Unauthorized access to the Wi-Fi credentials
        \\
        \hline       
        \textbf{Communication between device and its programmer}
        &
        \textcolor{magenta}{\textbf{Lack of strong mechanism to prevent replayed, old messages}} ($V_{SG}04$)
        &
        Used inexpensive hardware (like a USRP, DAQ, and antennas) to eavesdrop on the wireless transmission~\cite{2019survey_ref43}
        &
        Attacker within the device’s wireless range
        &
        Post-deployment phase
        &
        Implantable Cardiac Defibrillator
        &
        Battery drain attack by repeatedly replaying activation messages, session hijacking
        \\
        \hline
        \textbf{Firmware update channel}
        &
        \textcolor{magenta}{\textbf{Use of CRC-32 for integrity check, unencrypted firmware}} ($V_{SG}05$)
        &
        Intercepted firmware updates, reverse-engineered firmware, and bypassed CRC-32 via re-signed payload~\cite{2019survey_ref65}	
        &
        Control through an intermediary between the device and the cloud
        &
        Post-deployment phase
        &
        Withings Activite  (Wearable fitness device)
        &
        Unauthorized access, malicious firmware upload
        \\
        \hline
        \textbf{Communication between health tracker and its USB base}
        &
        \textcolor{magenta}{\textbf{Protocol configuration without encryption and authentication}} ($V_{SG}06$)
        &
        Retrieves data from the tracker, injects false values and uploads them into the account of the corresponding user on the web server~\cite{2019survey_ref66}
	      &
        Proximity-based wireless access (~15 ft range)
        &
        Post-deployment phase
        &
        Fitbit (Wearable fitness tracker)
        &
        Continuously send fake queries to the tracker device, rapidly draining its battery
        \\
        \hline
        
    \end{tabular}
    \end{adjustbox}
    
\end{table}

\setcounter{table}{4}

\begin{table}[H]
    \centering
    \caption{(Continued) Taxonomy of Attacks in IoMT: Between Sensor and Gateway Layer}
    \label{tab:tax_sensor2gateway_lyr}
    \renewcommand{\arraystretch}{1.5} 
    \rowcolors{2}{gray!10}{white}
    \begin{adjustbox}{width=1.2\textwidth,center=\textwidth}
    \begin{tabular}{>{\centering\arraybackslash}p{2.7cm}>{\centering\arraybackslash}p{2.7cm}>
    {\centering\arraybackslash}p{3.5cm}>
    {\centering\arraybackslash}p{1.7cm}>
    {\centering\arraybackslash}p{2cm}>
    {\centering\arraybackslash}p{2cm}>
    {\centering\arraybackslash}p{3.5cm}}
        \hline
        \rowcolor{gray!30}
        \textbf{Attack Surface} 
        & 
        \textbf{Vulnerability} ($V_{SG}$XX)
        & 
        \textbf{Attack Methodology}
        & 
        \textbf{Attacker's Access} 
        & 
        \textbf{Attack Stage} 
        & 
        \textbf{Target} 
        & 
        \textbf{Attack Impact} 
        \\
        \hline
        \textbf{Communication between fitness device and smartphone}
        &
        \textcolor{magenta}{\textbf{Long-term encryption key broadcasted in plaintext}} ($V_{SG}07$)
        &
        Researchers used Ubertooth, HCI snoop logs, and the Adafruit sniffer to capture and recover BLE traffic~\cite{2019survey_ref72}
        &
        Attacker is within the BLE range
        &
        Post-deployment, pairing phase
        &
        Amazon Amazfit (wearable fitness tracker)
        &
        Unauthorized access, decrypted communication
        \\
        \hline
        \textbf{Communication interface}
        &
        \textcolor{magenta}{\textbf{Continuous BLE advertising irrespective of whether the tracker is already paired}} ($V_{SG}08$); \textcolor{magenta}{\textbf{Use of fixed MAC address}} ($V_{SG}09$)
        &
        BlueZ (Linux Bluetooth stack) and GattTool utility were used to scan for BLE devices, check advertising behavior, flood the tracker with continuous read requests
        &
        Attacker within BLE range
        &
        Post-deployment phase
        &
        Fitbit Charge (Wearable fitness tracker)
        &
        MitM via MAC spoofing, tracking via static MAC, DoS attack, battery drain attack
        \\
        \hline
    \end{tabular}
    \end{adjustbox}
    
\end{table}

\subsection{Gateway Layer: Attack Surfaces and Vulnerabilities}
\begin{table}[H]
    \centering
    \caption{Taxonomy of Attacks in IoMT: Gateway Layer}
    \label{tab:tax_gateway_lyr}
    \renewcommand{\arraystretch}{1.5} 
    \rowcolors{2}{gray!10}{white}
    \begin{adjustbox}{width=1.2\textwidth,center=\textwidth}
    \begin{tabular}{>{\centering\arraybackslash}p{2.3cm}>{\centering\arraybackslash}p{2.5cm}>
    {\centering\arraybackslash}p{3cm}>
    {\centering\arraybackslash}p{3cm}>
    {\centering\arraybackslash}p{2cm}>
    {\centering\arraybackslash}p{2cm}>
    {\centering\arraybackslash}p{2.7cm}}
        \hline
        \rowcolor{gray!30}
        \textbf{Attack Surface} 
        & 
        \textbf{Vulnerability} ($V_G$XX)
        & 
        \textbf{Attack Methodology}
        & 
        \textbf{Attacker's Access} 
        & 
        \textbf{Attack Stage} 
        & 
        \textbf{Target} 
        & 
        \textbf{Impact of Attack} 
        \\
        \hline
        
        \textbf{Web management interface of medical gateway device}
        &
        \textcolor{magenta}{\textbf{Misfortune cookie vulnerability: No validation/limiting of data copied from cookie}} ($V_G01$)
        &
        Crafted HTTP cookie exploits the misfortune cookie flaw, allowing memory modification~\cite{2019survey_ref116}
        &
        Unauthenticated network access to the medical gateway
        &
        Post-deployment phase
        &
        Qualcomm Life Capsule's Datacaptor Terminal Server (DTS)
        &
        Attacker can disrupt communication between the hospital network and the connected bedside devices
        \\
        \hline

        \multirow{2}{*}{\textbf{Software}}
        &
        \textcolor{magenta}{\textbf{Unpatched software (e.g. use of insecure \nth{3} party libraries)}} ($V_G02$)
        &
        Network mapping and vulnerability assessments performed through Nmap and OpenVAS~\cite{2019survey_ref90}
        &
        Attacker can exploit known vulnerabilities in software
        &
        Post-deployment phase
        &
        Networked medical gateways
        &
        Gain unauthorized access, steal patient data, inject malware
        \\
        \cline{2-7}

        &
        \textcolor{magenta}{\textbf{Out-of-bounds write  (CWE-787)}} ($V_G03$)
        &
        No attack shown in CVE-2021-27410~\cite{CVE-2021-27410}
        &
        Not mentioned
        &
        Post-deployment phase
        &
        Welch Allyn Software Development Kit (SDK)
        &
        Corruption of data or code execution
        \\
        \hline

        \textbf{Network ports}
        &
        \textcolor{magenta}{\textbf{Open ports}} ($V_G04$); \textcolor{magenta}{\textbf{Default or weak passwords}} ($V_G05$)
        &
        Network mapping and vulnerability assessments performed through Nmap and OpenVAS~\cite{2019survey_ref90}
        &
        Remote attacker via open network ports
        &
        Post-deployment phase
        &
        Networked medical gateways
        &
        Unauthorized remote access leading to patient data theft, malware installations
        \\
        \hline

    \end{tabular}
    \end{adjustbox}
    
\end{table}

\setcounter{table}{5}

\begin{table}[H]
    \centering
    \caption{(Continued) Taxonomy of Attacks in IoMT: Gateway Layer}
    \label{tab:tax_gateway_lyr}
    \renewcommand{\arraystretch}{1.5} 
    \rowcolors{2}{gray!10}{white}
    \begin{adjustbox}{width=1.2\textwidth,center=\textwidth}
    \begin{tabular}{>{\centering\arraybackslash}p{2.3cm}>{\centering\arraybackslash}p{2.5cm}>
    {\centering\arraybackslash}p{3cm}>
    {\centering\arraybackslash}p{3cm}>
    {\centering\arraybackslash}p{2cm}>
    {\centering\arraybackslash}p{2cm}>
    {\centering\arraybackslash}p{2.7cm}}
        \hline
        \rowcolor{gray!30}
        \textbf{Attack Surface} 
        & 
        \textbf{Vulnerability} ($V_G$XX)
        & 
        \textbf{Attack Methodology}  
        & 
        \textbf{Attacker's Access} 
        & 
        \textbf{Attack Stage} 
        & 
        \textbf{Target} 
        & 
        \textbf{Attack Impact} 
        \\
        \hline

    \textbf{Stored data}
    &
    \textcolor{magenta}{\textbf{No encryption for sensitive data at rest}} ($V_G06$)
    &
    No attack shown in CVE-2019-18254~\cite{CVE-2019-18254}
    &
    Physical access to device
    &
    Post-deployment phase
    &
    BIOTRONIK CardioMessenger II
    &
    Unauthorized access to medical data
    \\
    \cline{2-7}
    &
    \textcolor{magenta}{\textbf{All user data, preferences, and sensor activity stored unencrypted in the gateway}} ($V_G07$)
    &
    Used reverse-engineering tools (e.g., APK Extractor, dex2jar) to decompile and analyse the app’s Java source code~\cite{2019survey_ref71}
    &
    Physical Access: Obtaining the gateway (android phone)
    &
    Post-deployment phase
    &
    Jawbone UP Move app in android phone
    &
    Violate data privacy
    \\
    \hline
    \multirow{2}{*}{\textbf{Gateway app}}
        &
        \textcolor{magenta}{\textbf{Firmware (binary APK file) is unencrypted}} ($V_G08$)
        &
        Make malicious changes to firmware, then push the modified  firmware to the fitness tracker~\cite{2019survey_ref69}
        &
        Ability to maliciously update the app in the gateway
        &
        Post-deployment phase
        &
        Gateway app of a fitness tracker 
        &
        Attacker can make malicious modification to the app’s functionality or to the stored firmware
    \\
    \cline{2-7}
    &
    \textcolor{magenta}{\textbf{App’s source code not obfuscated. Hence, easy to be reverse-
engineered.}} ($V_G09$)
    &
    Reverse engineer the app with JADX, disassemble and modify firmware via IDA Pro, then upload it to the fitness tracker~\cite{2019survey_ref69}
    &
    Access to reverse engineer and update the gateway app
    &
    Post-deployment phase
    &
    Gateway app of a fitness tracker (exact model not disclosed by the authors)
    &
    Attacker can make malicious modification to the app’s functionality or to the stored firmware
    \\
    \hline
        \textbf{USB ports and CD/DVD drives}
        &
        \textcolor{magenta}{\textbf{Unauthorized access to the storage port}} ($V_G10$)
        &
        Network mapping and vulnerability assessments performed through Nmap and OpenVAS~\cite{2019survey_ref90}
        &
        Physical access
        &
        Post-deployment phase
        &
        Networked medical gateways
        &
        Potential data privacy breach, malware installation
        \\
    \hline
    \multirow{2}{*}{\makecell{\textbf{Operating}\\ \textbf{System (OS)}}}
        &
        \textcolor{magenta}{\textbf{Privilege Escalation (CWE-269)}} ($V_G11$)
        &
        No attack shown in CVE-2021-32025~\cite{CVE-2021-32025}
        &
        Not mentioned
        &
        Post-deployment phase
        &
        QNX Neutrino Kernel
        &
        Unauthorized modification of settings, access sensitive data, or cause the system to crash
        \\
        \cline{2-7}
        &
        \textcolor{magenta}{\textbf{Outdated OS (e.g., Windows XP)}} ($V_G12$)
        &
        Network mapping and vulnerability assessments performed through Nmap and OpenVAS~\cite{2019survey_ref90}	
        &
        Attacker able to exploit known vulnerabilities in OS
        &
        Post-deployment phase
        &
        Networked medical gateways
        &
        Gain unauthorized access, steal patient data, inject malware
        \\
        \hline    

        \textbf{Antivirus}
        &
        \textcolor{magenta}{\textbf{Missing antivirus protection or outdated virus signatures}} ($V_G13$)
        &
        Network mapping and vulnerability assessments performed through Nmap and OpenVAS~\cite{2019survey_ref90}
        &
        Attacker can inject malware or virus in the medical device
        &
        Post-deployment (operational) phase
        &
        Networked medical gateways
        &
        DoS, ransomware
        \\
        \hline
    \end{tabular}
    \end{adjustbox}
    
\end{table}

\subsection{Between Gateway Layer and Cloud Layer: Attack Surfaces and Vulnerabilities}

\begin{table}[H]
    \centering
    \caption{Taxonomy of Attacks in IoMT: Between Gateway and Cloud Layer}
    \label{tab:tax_gateway2cloud_lyr}
    \renewcommand{\arraystretch}{1.5} 
    \rowcolors{2}{gray!10}{white}
    \begin{adjustbox}{width=1.2\textwidth,center=\textwidth}
    \begin{tabular}{>{\centering\arraybackslash}p{2.7cm}>{\centering\arraybackslash}p{2.5cm}>
    {\centering\arraybackslash}p{3.3cm}>
    {\centering\arraybackslash}p{1.7cm}>
    {\centering\arraybackslash}p{2cm}>
    {\centering\arraybackslash}p{2cm}>
    {\centering\arraybackslash}p{2.7cm}}
        \hline
        \rowcolor{gray!30}
        \textbf{Attack Surface} 
        & 
        \textbf{Vulnerability} ($V_{GC}$XX)
        & 
        \textbf{Attack Methodology}
        & 
        \textbf{Attacker's Access} 
        & 
        \textbf{Attack Stage} 
        & 
        \textbf{Target} 
        & 
        \textbf{Attack Impact} 
        \\
        \hline

        \textbf{Communication between medical device’s app and the server}
        &
        \textcolor{magenta}{\textbf{HTTP requests contain clear-text metadata with  sensitive information}} ($V_{GC}01$)
        &
        Inferred network traffic patterns and data~\cite{2019survey_ref114}
        &
        Attacker can observe the medical device’s Wi-Fi network
        &
        Post-deployment phase
        &
        Withings Blood Pressure Monitor
        &
        Attacker can infer user behaviour from metadata analysis
        \\
        \hline

        \textbf{DICOM communication with PACS server}
        &
        \textcolor{magenta}{\textbf{The DICOM standard supports encryption but is not enforced}} ($V_{GC}02$)
        &
        Network mapping and vulnerability assessments performed through Nmap and OpenVAS~\cite{2019survey_ref90}
        &
        Attacker can intercept  unencrypted DICOM images
        &
        Post-deployment phase
        &
        PACS server in hospitals
        &
        Unauthorized access to patient scans\\
        \hline

        \textbf{Communication between gateway and the web server}
        &
        \textcolor{magenta}{\textbf{Both login information and medical data are transmitted in cleartext form}} ($V_{GC}03$)
        &
        Discovers any Fitbit tracker device within a radius of 15 ft, injects false values and uploaded them into the account of the corresponding user on the web-server~\cite{2019survey_ref66}
        &
        Proximity-based wireless access (range: $\approx$ 15 ft)
        &
        Post-deployment phase
        &
        USB base (gateway) of wearable fitness tracker
        &
        Breach of private health data; forge activity data to earn financial rewards, false insurance claims
        \\
        \hline
        \textbf{Communication between fitness tracker’s android app and the web server}
        &
        \textcolor{magenta}{\textbf{Lack of robust certificate validation}} ($V_{GC}04$)
        &
        Set up proxy to intercept web traffic; bypass HTTPS encryption by installing a fake SSL Certificate in the android phone~\cite{2019survey_ref71}
        &
        Attacker able to install a malicious root CA certificate in the victim's phone
        &
        Post-deployment phase
        &
        Jawbone UP Move (fitness tracker) app in android phone (gateway)
        &
        Attacker can steal user’s credentials and activity data through a Man-in-the-Middle (MitM) attack
        \\
        \hline
        \textbf{Communication between fitness tracker’s android app and third party servers}
        &
        \textcolor{magenta}{\textbf{Extensive data sharing with third parties without user consent}} ($V_{GC}05$)
        &
        No attack exploiting this vulnerability was shown in~\cite{2019survey_ref71}
        &
        Attacker can eavesdrop the communication
        &
        Post-deployment phase
        &
        Jawbone UP Move (fitness tracker) app in android phone (gateway)
        &
        Risk of user-privacy breach by untrusted third parties
        \\
        \hline

    \end{tabular}
    \end{adjustbox}
    
\end{table}

\subsection{Cloud Layer: Attack Surfaces and Vulnerabilities}
\begin{table}[H]
    \centering
    \caption{Taxonomy of Attacks in IoMT: Cloud Layer}
    \label{tab:tax_cloud_lyr}
    \renewcommand{\arraystretch}{1} 
    \begin{adjustbox}{width=1.2\textwidth,center=\textwidth}
    \begin{tabular}{>{\centering\arraybackslash}p{2.3cm}>{\centering\arraybackslash}p{2.5cm}>
    {\centering\arraybackslash}p{2.2cm}>
    {\centering\arraybackslash}p{1.7cm}>
    {\centering\arraybackslash}p{2cm}>
    {\centering\arraybackslash}p{2cm}>
    {\centering\arraybackslash}p{2.7cm}}
        \hline
        \rowcolor{gray!30}
        \textbf{Attack Surface} 
        & 
        \textbf{Vulnerability} ($V_C$XX)
        & 
        \textbf{Attack Methodology} 
        & 
        \textbf{Attacker's Access} 
        & 
        \textbf{Attack Stage} 
        & 
        \textbf{Target} 
        & 
        \textbf{Attack Impact} 
        \\
        \hline
        \textbf{OS and software}
        &
        \textcolor{magenta}{\textbf{Outdated OS (e.g., Windows XP), unpatched software, insecure \nth{3} party libraries}} ($V_C01$)
        &
        Network mapping and vulnerability study performed through Nmap and OpenVAS~\cite{2019survey_ref90}
        &
        Attacker can exploit known vulnerabilities in OS or software
        &
        Post-deployment phase
        &
        PACS, EMR/EHR and other servers in hospitals
        &
        Unauthorized access, theft of patient data, malware injection
        \\
        \hline
        \textbf{Antivirus}, \textbf{Firewall}
        &
        \textcolor{magenta}{\textbf{Outdated antivirus protection and firewalls}} ($V_C02$)
        &
        Network mapping and vulnerability study performed through Nmap and OpenVAS~\cite{2019survey_ref90}
        &
        Attacker can inject malware remotely or through physical media
        &
        Post-deployment phase
        &
        PACS, EMR/EHR and other servers in hospitals
        &
        Unauthorized access to patient and hospital data
        \\
        \hline

        \textbf{Network ports}
        &
        \textcolor{magenta}{\textbf{Open ports}} ($V_C03$); \textcolor{magenta}{\textbf{Default or weak passwords}} ($V_C04$)
        &
        Network mapping and vulnerability study performed through Nmap and OpenVAS~\cite{2019survey_ref90}
        &
        Remote attacker via open network ports
        &
        Post-deployment phase
        &
        PACS, EMR/EHR and other servers in hospitals
        &
        Unauthorized remote access leading to patient data theft, malicious malware installations, mis-configured settings
        \\
        \hline

        \textbf{USB ports and CD/DVD drives}
        &
        \textcolor{magenta}{\textbf{Unrestricted storage access}} ($V_C05$)
        &
        Network mapping and vulnerability study performed through Nmap and OpenVAS~\cite{2019survey_ref90}
        &
        Attacker can physically plug in an infected, removable media into the target device
        &
        Post-deployment phase
        &
        PACS, EMR/EHR and other servers in hospitals
        &
        Insider attack:  Medical personnel can perform unauthorized copying of  patient data
        \\
        \hline        
        \textbf{VPN connections with medical vendor}
        &
        \textcolor{magenta}{\textbf{Vendors have unrestricted VPN access to the entire hospital network}} ($V_C06$)
        &
        Network mapping and vulnerability study performed through Nmap and OpenVAS shown in~\cite{2019survey_ref90}
        &
        Attacker can remote access hospital network via compromised VPN
        &
        Post-deployment phase; Network maintenance phase
        &
        PACS, EMR/EHR and other servers in hospitals
        &
        Unauthorized access to patient and hospital data
        \\
        \hline        
    \end{tabular}
    \end{adjustbox}
    
\end{table}

\setcounter{table}{7}

\begin{table}[H]
    \centering
    \caption{(Continued) Taxonomy of Attacks in IoMT: Cloud Layer}
    \label{tab:tax_cloud_lyr}
    \renewcommand{\arraystretch}{1.5} 
    \rowcolors{2}{gray!10}{white}
    \begin{adjustbox}{width=1.2\textwidth,center=\textwidth}
    \begin{tabular}{>{\centering\arraybackslash}p{2.3cm}>{\centering\arraybackslash}p{3.3cm}>
    {\centering\arraybackslash}p{2.2cm}>
    {\centering\arraybackslash}p{1.7cm}>
    {\centering\arraybackslash}p{2.6cm}>
    {\centering\arraybackslash}p{2cm}>
    {\centering\arraybackslash}p{2.8cm}}
        \hline
        \rowcolor{gray!30}
        \textbf{Attack Surface} 
        & 
        \textbf{Vulnerability} ($V_C$XX)
        & 
        \textbf{Attack Methodology}  
        & 
        \textbf{Attacker's Access} 
        & 
        \textbf{Attack Stage} 
        & 
        \textbf{Target} 
        & 
        \textbf{Attack Impact} 
        \\
        \hline
        \textbf{Data stored in server}
        &
        \textcolor{magenta}{\textbf{The terms claim all user data is encrypted, but the privacy policy admits server-stored data is not}} ($V_C07$)
        &
        No attack demonstrated in~\cite{2019survey_ref83}
        &
        Not applicable
        &
        Post-deployment (operational) stage
        &
        BASIS fitness tracker
        &
        Unauthorized access to user’s health and other data, loss of user trust in the fitness provide
        \\
        \cline{2-7}
        &
        \textcolor{magenta}{\textbf{Users can delete data from the device, but server-side deletion remains unclear}} ($V_C08$)
        &
        No attack demonstrated in~\cite{2019survey_ref83}
        &
        Not applicable
        &
        Post-account-termination
        &
        Fitbit fitness tracker
        &
        Same as the impacts of $V_C07$
        \\
        \hline
        \textbf{Digital medical records management software of EMR/EHR server}
        &
        \textcolor{magenta}{\textbf{Improper privilege management (CWE-269) allowing unauthorized access to the} $manage$\_$site$\_$files$ $.php$\textbf{ interface}} ($V_C09$)
        &
        No attack shown in CVE-2022-31496~\cite{CVE-2022-31496}
        &
        Not mentioned
        &
        Post-deployment phase
        &
        LibreHealth EHR Base
        &
        Unauthorized access to patient records
        \\
        \hline

        \textbf{Web application interface of medical records management software of EMR/EHR server}
        &
        \textcolor{magenta}{\textbf{Local File Inclusion (LFI) allowing inclusion and execution of arbitrary PHP files within the application}} ($V_C10$)
        &
        No attack shown in CVE-2020-11439~\cite{CVE-2020-11439}
        &
        Not mentioned
        &
        Post-deployment phase
        &
        LibreHealth EMR
        &
        Unauthorized access to sensitive data, malicious code injection
        \\
        \hline

        \textbf{Network user accounts}
        &
        \textcolor{magenta}{\textbf{Hardcoded/default password}} ($V_C11$)
        &
        No attack shown in CVE-2013-7442~\cite{CVE-2013-7442}
        &
        Not mentioned
        &
        Post-deployment phase
        &
        GE Healthcare Centricity PACS Workstation
        &
        Unauthorized access
        \\
        \hline
    \end{tabular}
    \end{adjustbox}
\end{table}

\subsection{Visualization Layer: Attack Surfaces and Vulnerabilities}

\begin{table}[H]
    \centering
    \caption{Taxonomy of Attacks in IoMT: Visualization Layer}
    \label{tab:tax_vis_lyr}
    \renewcommand{\arraystretch}{1.5} 
    \rowcolors{2}{gray!10}{white}
    \begin{adjustbox}{width=1.2\textwidth,center=\textwidth}
    \begin{tabular}{>{\centering\arraybackslash}p{2.3cm}>{\centering\arraybackslash}p{2.5cm}>
    {\centering\arraybackslash}p{2.2cm}>
    {\centering\arraybackslash}p{1.7cm}>
    {\centering\arraybackslash}p{1.5cm}>
    {\centering\arraybackslash}p{2cm}>
    {\centering\arraybackslash}p{2cm}}
        \hline
        \rowcolor{gray!30}
        \textbf{Attack Surface } 
        & 
        \textbf{Vulnerability} ($V_V$XX)
        & 
        \textbf{Attack Methodology} 
        & 
        \textbf{Attacker's Access} 
        & 
        \textbf{Attack Stage} 
        & 
        \textbf{Target} 
        & 
        \textbf{Attack Impact} 
        \\
        
        \hline
        \textbf{Informatics software for medical lab data management} 
        & 
        \textcolor{magenta}{\textbf{Insufficient session expiration (CWE-613)}} ($V_V01$)
        & 
        No attack shown in CVE-2022-30277~\cite{cve-2022-30277}
        & 
        Not mentioned
        & 
        Post-deployment phase
        & 
        BD Synapsys Informatics Solution
        & 
        Unauthorized access to sensitive information
        \\
        
        \hline
        \textbf{Central station in which the doctor views the status of multiple patients}
        & 
        \textcolor{magenta}{\textbf{Out-of-bounds write (CWE-787)}} ($V_V02$) 
        & 
        No attack shown in CVE-2021-27410~\cite{CVE-2021-27410}
        & 
        Not mentioned
        & 
        Post-deployment phase
        & 
        Welch Allyn Connex Central Station 
        & 
        Data corruption or malicious code execution \\
        
        \hline
    \end{tabular}
    \end{adjustbox}
    
\end{table}

\section{Translating the Proposed Attack Taxonomy into Actionable Security Measures}\label{sec:translation_countermeasures}

We use the insights from our attack taxonomy to establish a structured methodology that will enable security professionals to \textit{assess}, \textit{identify}, and \textit{mitigate} cybersecurity vulnerabilities in their IoMT systems. This methodology adopts a layer-wise approach in alignment with the proposed taxonomy. By methodically examining each layer, we provide a granular and targeted assessment of security risks, enabling engineers to implement layer-specific mitigation strategies that enhance the overall resilience of IoMT systems. 

\subsection{Security of Sensor/Actuator Layer}

\subsubsection*{Evaluate the physical security of the sensor-based medical device}
\begin{itemize}[leftmargin=*,noitemsep,parsep=0pt,partopsep=0pt]
    \item \textit{Vulnerability ($V_S01$ - Table~\ref{tab:tax_sensor_lyr}):} Device susceptibility to electro-magnetic interference (EMI).
    \item \textit{Recommended Security Measures:} Faraday shielding around sensing and pacing leads to block external electromagnetic signals; Twisted-pair wiring to reduce the susceptibility to EMI; Equip digital filters and error-checking mechanisms to reject false signals caused by EMI; Use Medical Implant Communication Service (MICS) band (402-405 MHz)~\cite{mics}, which has strict power limitations to reduce interference risks.
\end{itemize}

\subsubsection*{Check the external-media access points on the device}

\begin{itemize}[leftmargin=*,noitemsep,parsep=0pt,partopsep=0pt]
    \item \textit{Vulnerability ($V_S02$ - Table~\ref{tab:tax_sensor_lyr}):} Unrestricted access to physical storage interfaces such as USB ports and CD/DVD drives.
    \item \textit{Recommended Security Measures:} Disable USB ports and CD/DVD drives unless specifically needed; Restrict write permissions for USB ports.
\end{itemize}

\subsubsection*{Examine the device operator's environment for security risks}

\begin{itemize}[leftmargin=*,noitemsep,parsep=0pt,partopsep=0pt]
    \item \textit{Vulnerability ($V_S03$ - Table~\ref{tab:tax_sensor_lyr}):} Attackers can infer PINs or device commands by recording keypress vibrations and sounds from a smartwatch worn by the operator of a medical device.
    \item \textit{Recommended Security Measures:} Enforce hospital security policies to restrict personal smartwatches near medical devices; Prevent healthcare personals from allowing unnecessary access to their smartwatch motion sensors (accelerometer/gyroscope) and microphone by training them on cybersecurity risks; Replace numeric PINs with gesture-based or biometric authentication to eliminate keypad sounds; Randomize the positions of the keypad button on the medical device to prevent side-channel inference.
\end{itemize}

\subsubsection*{Analyze the security of the device software}
\begin{itemize}[leftmargin=*,noitemsep,parsep=0pt,partopsep=0pt]
    \item \textit{Vulnerability ($V_S04$ - Table~\ref{tab:tax_sensor_lyr}):} Unpatched or vulnerable software.
    \item \textit{Recommended Security Measures:} Use hardened containers (Docker~\cite{docker}, Windows Sandbox~\cite{sandbox}, Kubernetes~\cite{kubernetes}) to isolate outdated, vulnerable applications from other resources in the medical network; Use cryptographic signatures to verify (app) binary authenticity.
\end{itemize}

\begin{itemize}[leftmargin=*,noitemsep,parsep=0pt,partopsep=0pt]
    \item \textit{Vulnerability ($V_S05$; $V_S06$ - Table~\ref{tab:tax_sensor_lyr}):} Out-of-bounds write; Buffer overflow.
    \item \textit{Recommended Security Measures:} Avoid applications written using memory-unsafe programming languages (e.g., C, C++); Apply schemes for protection against buffer-overflow and out-of-bounds write in firmware updates; Perform static code analysis and dynamic fuzz testing (using automated tools like AFL~\cite{afl}, libFuzzer~\cite{libfuzzer}) to identify out-of-bounds vulnerabilities before deploying medical software; Use hardware-based protections like ARM TrustZone~\cite{armtrustzone} or Intel Memory Protection Extensions (MPX)~\cite{intelmpx} that can detect and prevent buffer overflow exploits.
\end{itemize}

\subsubsection*{Examine the user access to the device}
\begin{itemize}[leftmargin=*,noitemsep,parsep=0pt,partopsep=0pt]
    \item \textit{Vulnerability ($V_S07$; $V_S12$ - Table~\ref{tab:tax_sensor_lyr}):} Use of hard-coded passwords; Default or weak passwords.
    \item \textit{Recommended Security Measures:} Replace hard-coded or default credentials with biometric or OTP-based authentication; Prevent brute-force attacks by implementing account lockouts after multiple failed login attempts; Ensure that all devices require password changes upon deployment and disallow hard-coded credentials in firmware; Use complex alphanumeric passwords with regular rotation enforced by the hospital’s device management system.
\end{itemize}

\subsubsection*{Examine the configuration file(s) in the device}
\begin{itemize}[leftmargin=*,noitemsep,parsep=0pt,partopsep=0pt]
    \item \textit{Vulnerability ($V_S08$ - Table~\ref{tab:tax_sensor_lyr}):}  Configuration files with plaintext passwords.
    \item \textit{Recommended Security Measures:} Use PBKDF2, bcrypt, or Argon2~\cite{medium} for secure password hashing instead of plaintext storage; Store encrypted credentials in tamper-proof hardware (e.g., TPM~\cite{tpm} or Secure Enclave~\cite{secureenclave}) instead of software configuration files; Use Role-Based Access Control (RBAC):  Restrict access to critical files and resources to authorized users (e.g., hospital IT personnel) only; Deploy Multi-Factor Authentication (MFA) for access to critical files and resources.
\end{itemize}

\subsubsection*{Evaluate the security of Operating System (OS)}

\begin{itemize}[leftmargin=*,noitemsep,parsep=0pt,partopsep=0pt]
    \item \textit{Vulnerability ($V_S09$ - Table~\ref{tab:tax_sensor_lyr}):} Privilege escalation.
    \item \textit{Recommended Security Measures:} Enforce role-based access control (RBAC) and least privileges for both users and applications; Verify digital signatures for all firmware and OS components before execution; Apply MAC frameworks like SELinux~\cite{selinux} or AppArmor~\cite{apparmor} to isolate processes and prevent unauthorized privilege elevation.
\end{itemize}

\begin{itemize}[leftmargin=*,noitemsep,parsep=0pt,partopsep=0pt]
    \item \textit{Vulnerability ($V_S10$ - Table~\ref{tab:tax_sensor_lyr}):} Outdated or legacy OS.
    \item \textit{Recommended Security Measures:} Migrate from legacy OS to secure, modern operating systems with long-term support. For example, Windows 10/11 Long-Term Servicing Channel (LTSC) has the option of paid Extended Security Updates (ESU)~\cite{esu} for critical industries like healthcare; Run legacy applications in a secure virtualized environment on a modern OS. For example, if a radiology workstation still depends on Windows XP OS, instead of running XP directly on the CPU, hospitals can run it as a sandboxed (isolated) system within a Windows 10 Hyper-V~\cite{hyperv} virtual machine.
\end{itemize}

\subsubsection*{Assess the network ports}
\begin{itemize}[leftmargin=*,noitemsep,parsep=0pt,partopsep=0pt]
    \item \textit{Vulnerability ($V_S11$ - Table~\ref{tab:tax_sensor_lyr}):} Open network ports.
    \item \textit{Recommended Security Measures:} Disable unused or outdated services (e.g., Telnet~\cite{telnet}, FTP~\cite{ftp}); Deploy intrusion detection systems (IDS) or firewalls to monitor network traffic and block unauthorized access to ports; Regularly scan for open ports and enforce network segmentation to minimize exposure.
\end{itemize}

\subsubsection*{Check the antivirus}
\begin{itemize}[leftmargin=*,noitemsep,parsep=0pt,partopsep=0pt]
    \item \textit{Vulnerability ($V_S13$ - Table~\ref{tab:tax_sensor_lyr}):} 
         Missing antivirus.
    \item \textit{Recommended Security Measures:}
            Deploy lightweight antivirus solutions (e.g., McAfee Embedded Control~\cite{mcafee}, Windows Defender ATP for IoT~\cite{defender}); Complement endpoint defense with network-based IDS to detect malware propagation
\end{itemize}

\subsubsection*{Examine the security of device firmware}

\begin{itemize}[leftmargin=*,noitemsep,parsep=0pt,partopsep=0pt]
    \item \textit{Vulnerability ($V_S14$ - Table~\ref{tab:tax_sensor_lyr}):} Lack of protection against direct read/write access to firmware memory.
    \item \textit{Recommended Security Measures:} Store critical firmware components in protected, non-writable, read-only sections (Read-Only Memory (ROM)) of memory; Use Secure Enclaves or Memory Protection Units (MPUs) to store firmware in a tamper-resistant environment; Encrypt firmware not just in transit (during updates) but also at rest; Store cryptographic keys in tamper-proof hardware; Enforce secure boot to ensure only authenticated firmware runs.
\end{itemize}

\begin{itemize}[leftmargin=*,noitemsep,parsep=0pt,partopsep=0pt]
    \item \textit{Vulnerability ($V_S15$; $V_S16$ - Table~\ref{tab:tax_sensor_lyr}):} No firmware signature verification; Firmware integrity check relies on weak CRC values.
    \item \textit{Recommended Security Measures:} Replace weak CRC checks with cryptographic hash functions (e.g. SHA-256) to verify firmware integrity before every \textit{installation} and \textit{execution}; Design devices with rollback protection to prevent downgrades to vulnerable firmware versions. For example, the medical device can store the firmware version number in secure storage (e.g., Trusted Platform Module (TPM)~\cite{tpm}, or Secure Enclave~\cite{secureenclave}) and reject any downgrade attempt.
\end{itemize}


\subsubsection*{Inspect the network communication interface of the device}

\begin{itemize}[leftmargin=*,noitemsep,parsep=0pt,partopsep=0pt]
    \item \textit{Vulnerability ($V_S17$ - Table~\ref{tab:tax_sensor_lyr}):} 
         Presence of a factory account with a hard-coded password. 
    \item \textit{Recommended Security Measures:}
            Implement unique per device passwords instead of factory-set default passwords; After first-time setup, force the user to set a new password; Factory accounts should be removed or disabled before deployment; Also, refer to the security measures given for $V_S07$ and $V_S12$.
\end{itemize}

\begin{itemize}[leftmargin=*,noitemsep,parsep=0pt,partopsep=0pt]
    \item \textit{Vulnerability ($V_S18$; $V_S19$ - Table~\ref{tab:tax_sensor_lyr}):} Improper Access Control; Lack of certificate validation.
    \item \textit{Recommended Security Measures:} FTP is outdated and should be replaced with secure alternatives (e.g., SSH File Transfer Protocol (SFTP)~\cite{sftp}). If FTP is necessary, require certificate-based authentication; Implement Access Control Lists (ACLs): restrict which IPs or users can access the FTP service; Enforce network segmentation; Validate entire certificate chain; Enable certificate pinning to prevent MitM attack.
\end{itemize}

\subsection{Security of Communication between the Sensor/Actuator Layer and the Gateway Layer}

\subsubsection*{Examine the security of IoMT communication protocols}
\begin{itemize}[leftmargin=*,noitemsep,parsep=0pt,partopsep=0pt]
    \item \textit{Vulnerability ($V_{SG}01$ - Table~\ref{tab:tax_sensor2gateway_lyr}):} The DICOM standard supports encryption but is not enforced in the DICOM communication from imaging devices.
    \item \textit{Recommended Security Measures:} Ensure encryption of DICOM images and metadata using standardized protocols (e.g., TLS 1.3) before transmission; Implement IPsec VPN tunnels to secure image transmission between imaging equipment and PACS servers; Configure systems to reject unencrypted DICOM associations.
\end{itemize}

\begin{itemize}[leftmargin=*,noitemsep,parsep=0pt,partopsep=0pt]
    \item \textit{Vulnerability ($V_{SG}02$; $V_{SG}03$ - Table~\ref{tab:tax_sensor2gateway_lyr}):} Weak Wi-Fi Security such as the use of WPS PIN that can be discovered by a brute-force attack; Wi-Fi communication based on the 802.11 standard which can be attacked by flooding with spoofed deauthentication packets.
    \item \textit{Recommended Security Measures:} Use WPA3-Enterprise~\cite{wpa3e} with 802.1X authentication; Use Management Frame Protection (MFP)~\cite{mfp} to prevent de-authentication flooding attacks; Use MAC filtering as a secondary control and monitor Wi-Fi networks for intrusion attempts.
\end{itemize}

\subsubsection*{Examine the integrity and authenticity of communication protocols}
\begin{itemize}[leftmargin=*,noitemsep,parsep=0pt,partopsep=0pt]
    \item \textit{Vulnerability ($V_{SG}04$ - Table~\ref{tab:tax_sensor2gateway_lyr}):} Lack of strong mechanism to prevent replayed, old messages.
    \item \textit{Recommended Security Measures:} Introduce time-stamping mechanisms to detect and reject old messages; Enforce strict session expiry using session-based encryption keys; Implement nonce-based authentication to ensure each session has a unique, one-time-use cryptographic token; Implement strict API access controls to limit unauthorized queries.
\end{itemize}

\begin{itemize}[leftmargin=*,noitemsep,parsep=0pt,partopsep=0pt]
    \item \textit{Vulnerability ($V_{SG}05$; $V_{SG}06$ - Table~\ref{tab:tax_sensor2gateway_lyr}):} 
         Lack of proper integrity checks and encryption during firmware updates; Protocol configuration without encryption and authentication.
    \item \textit{Recommended Security Measures:}
            Encrypt firmware during updates using end-to-end encryption (E2EE); Use cryptographic hash functions (e.g., SHA-256) to digitally sign firmware updates for preventing unauthorized tampering.
\end{itemize}

\subsubsection*{Examine the implementation of (secure)  communication protocols}

\begin{itemize}[leftmargin=*,noitemsep,parsep=0pt,partopsep=0pt]
    \item \textit{Vulnerability ($V_{SG}07$ - Table~\ref{tab:tax_sensor2gateway_lyr}):} 
        Long-term encryption key broadcasted in plaintext.
    \item \textit{Recommended Security Measures:}
            Enable LE Secure Connections (Bluetooth 4.2 and later); Disable legacy pairing modes; Avoid broadcasting long-term keys by using ephemeral session keys with proper key rotation; Enforce authenticated and encrypted communication between Bluetooth devices.
\end{itemize}

\begin{itemize}[leftmargin=*,noitemsep,parsep=0pt,partopsep=0pt]
    \item \textit{Vulnerability ($V_{SG}08$; $V_{SG}09$ - Table~\ref{tab:tax_sensor2gateway_lyr}):} Continuous BLE advertising; Fixed MAC addresses.
    \item \textit{Recommended Security Measures:} Enable BLE privacy extensions to randomize MAC addresses periodically; Use adaptive BLE scanning to reduce BLE advertisement intervals or completely stop advertising once paired; Enforce user-consent mechanisms before a device starts BLE broadcasting.
\end{itemize}

\subsection{Security of Gateway Layer}

\subsubsection*{Examine the web interface of the gateway}
\begin{itemize}[leftmargin=*,noitemsep,parsep=0pt,partopsep=0pt]
    \item \textit{Vulnerability ($V_G$01 - Table~\ref{tab:tax_gateway_lyr}):}  Misfortune cookie vulnerability.
    \item \textit{Recommended Security Measures:} Patch vulnerable web server components (e.g., RomPager); Enforce session expiration to ensure cookies do not persist indefinitely; Use Secure, HttpOnly, and SameSite cookies to prevent modification by attackers~\cite{cookies}; Disable web interface access over public networks.
\end{itemize}

\subsubsection*{Examine the software(s) installed in the gateway}
\begin{itemize}[leftmargin=*,noitemsep,parsep=0pt,partopsep=0pt]
    \item \textit{Vulnerability ($V_G02$; $V_G03$ - Table~\ref{tab:tax_gateway_lyr}):} Unpatched software; Out-of-bounds write
    \item \textit{Recommended Security Measures:} Those recommended for mitigating the same vulnerabilities ($V_S04$; $V_S05$ - Table~\ref{tab:tax_sensor_lyr}) in the sensor/actuator layer.
\end{itemize}

\subsubsection*{Examine the network ports of the gateway}
\begin{itemize}[leftmargin=*,noitemsep,parsep=0pt,partopsep=0pt]
    \item \textit{Vulnerability ($V_G04$; $V_G05$ - Table~\ref{tab:tax_gateway_lyr}):} Open network ports; Default or weak passwords.
    \item \textit{Recommended Security Measures:} Those recommended for mitigating the same vulnerabilities ($V_S11$; $V_S12$ - Table~\ref{tab:tax_sensor_lyr}) in the sensor/actuator layer.
\end{itemize}

\subsubsection*{Examine the data stored in the gateway}
\begin{itemize}[leftmargin=*,noitemsep,parsep=0pt,partopsep=0pt]
    \item \textit{Vulnerability ($V_G06$; $V_G07$ - Table~\ref{tab:tax_gateway_lyr}):} Lack of encryption for sensitive data at rest; All user data, preferences, and sensor activity stored unencrypted in the gateway.
    \item \textit{Recommended Security Measures:} Store minimal patient data locally and prefer cloud-based access with strong encryption; Implement standardized encryption; Use Hardware Security Modules (HSMs) for storing encryption keys securely; Use file system encryption (e.g., LUKS~\cite{luks}, BitLocker~\cite{bitlocker}) for local storage. 
\end{itemize}

\subsubsection*{Examine the gateway app}

\begin{itemize}[leftmargin=*,noitemsep,parsep=0pt,partopsep=0pt]
    \item \textit{Vulnerability ($V_G08$ - Table~\ref{tab:tax_gateway_lyr}):} Latest firmware (binary) stored unencrypted in a directory directory of the app’s APK file.
    \item \textit{Recommended Security Measures:} Encrypt firmware binaries before embedding in the APK; Apply strict file access control and use Android keystore to protect decryption keys; Implement biometric authentication for accessing sensitive app-files; Use file system encryption for local storage.
\end{itemize}

\begin{itemize}[leftmargin=*,noitemsep,parsep=0pt,partopsep=0pt]
    \item \textit{Vulnerability ($V_G09$ - Table~\ref{tab:tax_gateway_lyr}):} App’s source code not obfuscated. Hence, it is easy to be reverse engineered.
    \item \textit{Recommended Security Measures:} Use code obfuscation tools (e.g., ProGuard~\cite{proguard}, R8~\cite{r8}, DexGuard~\cite{dexguard}) before deployment to prevent attackers from extracting app logic; Ensure debug symbols are removed before deployment; Encrypt and store app's compiled binaries; Use cryptographic signatures to verify (app) binary authenticity and detect unauthorized modifications (post reverse-engineering).
\end{itemize}

\subsubsection*{Check the external-media access points on the gateway}
\begin{itemize}[leftmargin=*,noitemsep,parsep=0pt,partopsep=0pt]
    \item \textit{Vulnerability ($V_G10$ - Table~\ref{tab:tax_gateway_lyr}):} Anyone can plug in an external storage device (like a USB flash drive or an external hard disk) or insert a CD/DVD into a system without any restriction.
    \item \textit{Recommended Security Measures:} Use BIOS/UEFI security settings to disable unauthorized USB device connections; Those recommended for mitigating the same vulnerability ($V_S02$ - Table~\ref{tab:tax_sensor_lyr}) in the sensor/actuator layer.
\end{itemize}

\subsubsection*{Check the security of the gateway OS}
\begin{itemize}[leftmargin=*,noitemsep,parsep=0pt,partopsep=0pt]
    \item \textit{Vulnerability ($V_G11$; $V_G12$ - Table~\ref{tab:tax_gateway_lyr}):} Privilege escalation; Outdated or legacy OS.
    \item \textit{Recommended Security Measures:} Those recommended for mitigating the same vulnerabilities ($V_S09$; $V_S10$ - Table~\ref{tab:tax_sensor_lyr}) in the sensor/actuator layer.
\end{itemize}

\subsubsection*{Check the antivirus in the gateway}
\begin{itemize}[leftmargin=*,noitemsep,parsep=0pt,partopsep=0pt]
    \item \textit{Vulnerability ($V_G13$ - Table~\ref{tab:tax_gateway_lyr}):} 
        Missing anti-virus protection or outdated virus signatures.
    \item \textit{Recommended Security Measures:}
            Those recommended for mitigating the same vulnerability ($V_S13$ - Table~\ref{tab:tax_sensor_lyr}) in the sensor/actuator layer.
\end{itemize}

\subsection{Security of Communication between the Gateway and the Cloud Layer}

\subsubsection*{Examine the security of communication protocols}
\begin{itemize}[leftmargin=*,noitemsep,parsep=0pt,partopsep=0pt]
    \item \textit{Vulnerability ($V_{GC}01$ - Table~\ref{tab:tax_gateway2cloud_lyr}):} HTTP requests sent by the medical device’s app (in smartphone) contain clear-text metadata. 
    \item \textit{Recommended Security Measures:} Replace clear-text metadata with random, non-reversible tokens; Enforce HTTPS with TLS 1.3 encryption for all data transmissions; Enable HSTS (HTTP Strict Transport Security)~\cite{hsts}; Route data through an IPsec or WireGuard VPN.
\end{itemize}

\subsubsection*{Examine the implementation of (secure) communication protocols}
\begin{itemize}[leftmargin=*,noitemsep,parsep=0pt,partopsep=0pt]
    \item \textit{Vulnerability ($V_{GC}02$ - Table~\ref{tab:tax_gateway2cloud_lyr}):} The DICOM standard supports encryption but is not enforced.
    \item \textit{Recommended Security Measures:} Use DICOM over TLS (DICOMweb~\cite{dicomweb}) instead of unencrypted DICOM transfers; Those recommended for mitigating the same vulnerability ($V_{SG}01$ - Table~\ref{tab:tax_sensor2gateway_lyr}) in the communication between the sensor/actuator layer and the gateway layer.
\end{itemize}

\begin{itemize}[leftmargin=*,noitemsep,parsep=0pt,partopsep=0pt]
    \item \textit{Vulnerability ($V_{GC}03$ - Table~\ref{tab:tax_gateway2cloud_lyr}):} Both login information and fitness data are transmitted in cleartext form.
    \item \textit{Recommended Security Measures:} Encrypt data in transit with TLS 1.3; Replace basic authentication (username/password over HTTP) with OAuth 2.0 token-based authentication; Use short-lived access tokens and refresh tokens to minimize exposure.
\end{itemize}

\begin{itemize}[leftmargin=*,noitemsep,parsep=0pt,partopsep=0pt]
    \item \textit{Vulnerability ($V_{GC}04$ - Table~\ref{tab:tax_gateway2cloud_lyr}):} Lack of robust certificate validation (only checking CA signatures but not Common Name (CN)).
    \item \textit{Recommended Security Measures:} Ensure both CN and Subject Alternative Name (SAN) are validated in all SSL/TLS certificates; Implement certificate pinning; Use SSL/TLS monitoring tools (e.g., Zeek~\cite{zeek}, Wireshark~\cite{wireshark}, ZAP~\cite{zap}) to detect unusual handshake patterns. Automatically reject expired or self-signed certificates.
\end{itemize}

\begin{itemize}[leftmargin=*,noitemsep,parsep=0pt,partopsep=0pt]
    \item \textit{Vulnerability ($V_{GC}05$ - Table~\ref{tab:tax_gateway2cloud_lyr}):} Data sharing with third parties without user consent.
    \item \textit{Recommended Security Measures:} Provide users with explicit opt-in/opt-out options for data sharing; Allow users to opt-out of non-essential data sharing via privacy settings; Ensure all third-party services comply with HIPAA~\cite{hipaa} and GDPR~\cite{gdpr} regulations; Replace Personally Identifiable Information (PII) with tokens before sharing with third-party data analytics.
\end{itemize}

\subsection{Security of Cloud Layer}

\subsubsection*{Evaluate the server OS, software applications for security risks}
\begin{itemize}[leftmargin=*,noitemsep,parsep=0pt,partopsep=0pt]
    \item \textit{Vulnerability ($V_C01$ - Table~\ref{tab:tax_cloud_lyr}):} Outdated OS
(e.g., Windows XP), unpatched software, insecure $3^{rd}$
party libraries.
    \item \textit{Recommended Security Measures:} Those recommended for mitigating  (similar) vulnerabilities $V_S04$, $V_S05$, $V_S06$, $V_S09$, $V_S10$ (Table~\ref{tab:tax_sensor_lyr}) in the sensor/actuator layer.
\end{itemize}

\subsubsection*{Evaluate the  antivirus and/or firewall protection}
\begin{itemize}[leftmargin=*,noitemsep,parsep=0pt,partopsep=0pt]
    \item \textit{Vulnerability ($V_C02$ - Table~\ref{tab:tax_cloud_lyr}):} 
        Lack of or outdated antivirus protection and firewalls.
    \item \textit{Recommended Security Measures:}
            Those recommended for mitigating the same vulnerability ($V_S13$ - Table~\ref{tab:tax_sensor_lyr}) in the sensor/actuator layer.
\end{itemize}

\subsubsection*{Evaluate the  security of the network ports and media access points of the server}
\begin{itemize}[leftmargin=*,noitemsep,parsep=0pt,partopsep=0pt]
    \item \textit{Vulnerability ($V_C03$; $V_C04$ - Table~\ref{tab:tax_gateway_lyr}):} Open network ports; Default or weak passwords.
    \item \textit{Recommended Security Measures:} Those recommended for mitigating the same vulnerabilities ($V_S11$; $V_S12$ - Table~\ref{tab:tax_sensor_lyr}) in the sensor/actuator layer.
\end{itemize}

\begin{itemize}[leftmargin=*,noitemsep,parsep=0pt,partopsep=0pt]
    \item \textit{Vulnerability ($V_C05$ - Table~\ref{tab:tax_cloud_lyr}):} Unauthorized access through an external storage device (like a USB flash drive or an external hard disk) or insert a CD/DVD into a system without any restriction.
    \item \textit{Recommended Security Measures:} Track USB insertions, activities, and removals using SIEM tools~\cite{siem}; Countermeasures recommended for mitigating (similar) vulnerabilities $V_S02$ (Table~\ref{tab:tax_sensor_lyr}) and $V_G10$ (Table~\ref{tab:tax_gateway_lyr}) in the sensor/actuator and the gateway layers, respectively.
\end{itemize}

\subsubsection*{Check the VPNs that access the server} 
\begin{itemize}[leftmargin=*,noitemsep,parsep=0pt,partopsep=0pt]
    \item \textit{Vulnerability ($V_C06$ - Table~\ref{tab:tax_cloud_lyr}):} Unrestricted VPN access to the entire hospital network leads to compromised VPN and infiltrate in the hospital’s network.
    \item \textit{Recommended Security Measures:} Restrict vendor VPN access to only required systems, not the entire network. Use Least Privilege Access (LPA) to grant minimal access to external vendors; Set VPN session timeouts (e.g., auto-disconnect after 30 minutes of inactivity)
\end{itemize}

\subsubsection*{Review the stored data for security and privacy risks}

\begin{itemize}[leftmargin=*,noitemsep,parsep=0pt,partopsep=0pt]
    \item \textit{Vulnerability ($V_C07$ - Table~\ref{tab:tax_cloud_lyr}):} The terms claim all user data is encrypted and confidential, but the privacy policy states that data in the database (server) is not encrypted.
    \item \textit{Recommended Security Measures:} Encrypt all stored health records, DICOM images, and personal data with standardized encryption (e.g., AES-256); Use GDPR-compliant data retention and deletion policies.
\end{itemize}

\begin{itemize}[leftmargin=*,noitemsep,parsep=0pt,partopsep=0pt]
    \item \textit{Vulnerability ($V_C08$ - Table~\ref{tab:tax_cloud_lyr}):} User can delete activity and sleep data from the device, but it is unclear whether all the user’s data stored in the servers are also erased. After account termination, personally identifiable data is removed, but de-identified historical data may still be used.
    \item \textit{Recommended Security Measures:} Use GDPR-compliant data retention and deletion policies; Automate regular auditing of server databases for obsolete patient data (e.g., post account termination) and their secure removal; Implement log monitoring tools to track data deletion events; Track all data deletion actions to ensure compliance and prevent unauthorized recovery.
\end{itemize}

\subsubsection*{Examine the digital medical records management software of EMR/EHR server}
\begin{itemize}[leftmargin=*,noitemsep,parsep=0pt,partopsep=0pt]
    \item \textit{Vulnerability ($V_C09$ - Table~\ref{tab:tax_cloud_lyr}):} Improper privilege management (CWE-269) allowing unauthorized access.
    \item \textit{Recommended Security Measures:} Monitor privileged user sessions using session recording tools (e.g., Teramind, ObserveIT); Countermeasures recommended for mitigating vulnerability $V_S09$ (Table~\ref{tab:tax_sensor_lyr}) in the sensor/actuator layer.
\end{itemize}

\subsubsection*{Examine the PHP scripts running on the server for database interaction}
\begin{itemize}[leftmargin=*,noitemsep,parsep=0pt,partopsep=0pt]
    \item \textit{Vulnerability ($V_C10$ - Table~\ref{tab:tax_cloud_lyr}):}  Improper input validation: Local File Inclusion (LFI) allowing inclusion and execution of arbitrary PHP files within the application.
    \item \textit{Recommended Security Measures:} Prevent attackers from including external or local files remotely (e.g., by setting $allow\_url\_include$ = `Off' and $allow\_url\_fopen$ = `Off' in PHP configuration); Only allow specific, predefined (full) file-paths (whitelist); Use Web Application Firewalls (WAFs) (e.g., ModSecurity~\cite{modsecurity}, Cloudflare~\cite{cloud}) to filter Local File Inclusion (LFI) attack patterns; Conduct penetration testing (pentesting) on cloud-based applications before release.
\end{itemize}

In addition to the above measures for LFI attacks, we recommend Table~\ref{tab:preventive_measures} that contains measures to prevent some common attacks.

\subsubsection*{Examine user access to the server}
\begin{itemize}[leftmargin=*,noitemsep,parsep=0pt,partopsep=0pt]
    \item \textit{Vulnerability ($V_C11$ - Table~\ref{tab:tax_cloud_lyr}):} (CVE-2013-
7442) The system uses the password `CANal1’ for the Administrator user and `iis’ for the IIS user. NOTE: it is not clear whether this password is default, hardcoded, or dependent on another system or product that requires a fixed value.
    \item \textit{Recommended Security Measures:} Refer to the security measures given for vulnerability $V_S17$ (Table~\ref{tab:tax_sensor_lyr}) in the sensor/actuator layer.
\end{itemize}

\subsection{Security of the Visualization Layer}

\subsubsection*{Examine the informatics software for medical lab data management integrating multiple lab equipment}
\begin{itemize}[leftmargin=*,noitemsep,parsep=0pt,partopsep=0pt]
    \item \textit{Vulnerability ($V_V01$ - Table~\ref{tab:tax_vis_lyr}):} 
    Insufficient session expiration (CWE-613): If a user forgets to log out or closes their browser, an attacker might be able to reopen the session and access sensitive data.
    \item \textit{Recommended Security Measures:} Enforce automatic logout after inactivity; Warn users if closing the browser without logout; Require Multi-Factor Authentication (e.g., SMS OTP, authenticator apps, biometric authentication) for re-login; Restrict sessions to one active login at a time (especially for admin/doctor roles); Allow dashboard access only from hospital-approved devices~\cite{ztna}.
\end{itemize}

\subsubsection*{Examine the central station that allows the doctor to view the status of multiple patients}
\begin{itemize}[leftmargin=*,noitemsep,parsep=0pt,partopsep=0pt]
    \item \textit{Vulnerability ($V_V02$ - Table~\ref{tab:tax_vis_lyr}):} Out-of-bounds write.
    \item \textit{Recommended Security Measures:} Those recommended for mitigating the same vulnerability ($V_S05$ - Table~\ref{tab:tax_sensor_lyr}) in the sensor/actuator layer.
\end{itemize}

\begin{table}[H]
    \centering
    \caption{Preventive Measures Against Common Web Attacks}
    \label{tab:preventive_measures}
    \renewcommand{\arraystretch}{1.5} 
    \rowcolors{2}{gray!10}{white}
    \begin{adjustbox}{width=1.2\textwidth,center=\textwidth}
    \begin{tabular}{>{\centering\arraybackslash}p{3.5cm}>{\centering\arraybackslash}p{3.5cm}>{\centering\arraybackslash}p{6cm}}
        \hline
        \rowcolor{gray!30}
        \textbf{Preventive Measure} 
        & 
        \textbf{Attack} 
        & 
        \textbf{What the Attack Does} 
        \\
        \hline
        \textbf{Use \texttt{Prepared} Statements} 
        & 
        SQL Injection (SQLi) 
        & 
        Injects malicious SQL to access or modify database records. 
        \\
        \textbf{Use \texttt{htmlspecialchars()} to escape user input} 
        & 
        Cross-Site Scripting (XSS) 
        & 
        Inserts malicious JavaScript to steal cookies, hijack sessions, or manipulate webpage content. 
        \\
        \textbf{Use \texttt{escapeshellarg()} to sanitize input} 
        & 
        Remote Code Execution (RCE) 
        & 
        Executes unauthorized commands on the server. 
        \\
        \hline
    \end{tabular}
    \end{adjustbox}
\end{table}

\section{Standards and Compliance}
Due to the critical interplay between clinical safety, system safety, and cybersecurity, IoMT components must adhere to strict regulatory compliance. Traditionally, medical devices were only subjected to safety and clinical compliance studies. In recent times, the threat of cyberattacks has led to the inclusion of security-specific checks in those compliance. For example, the Medical Device Coordination Group (MDCG) in the European Union released a specific note in 2019 focused on cybersecurity of medical devices~\cite{mdcg_europe_cybersecurity}. In 2023, the US Food and Drug Administration issued guidelines for the cybersecurity of medical devices~\cite{US_FDA_Cyber}. In 2024, a cybersecurity labeling scheme for medical devices was introduced in Singapore~\cite{Singapore_MD_Cyber}. The standardization flows, along with the cybersecurity implications, differ significantly depending on the device classification, which we discuss next.

\subsection{Medical Device Classification}
Although the classification of medical devices varies between countries, the underlying framework is largely consistent, with the risk level serving as the primary basis for the classification. For example, Singapore's Health Sciences Authority (HSA) categorizes medical devices into two broad segments: general medical devices and in-vitro diagnostic (IVD) devices. Within each segment, devices are further classified according to their associated risk levels, as shown in Tables~\ref{tab:risk_class_medical} and~\ref{tab:risk_class_ivd_medical}.

\begin{table}[hbt]
 \centering
 \caption{Risk classification of medical devices according to HSA, Singapore.}
 \label{tab:risk_class_medical}
 \begin{tabular}{c|c|c} \hline
 \rowcolor{gray!30} \textbf{CLASS} & \textbf{RISK LEVEL} & \textbf{EXAMPLES} \\ \hline
 A & Low Risk & Wheelchairs, Tongue depressors \\ \hline
 B & Low-moderate Risk & Hypodermic needles, Suction equipment \\ \hline
 C & Moderate-high Risk & Ventilators, Bone fixation plates \\ \hline
 D & High Risk & Heart valves, Implantable defibrillators \\ \hline
 \end{tabular}
\end{table}

\begin{table}[hbt]
\centering
\caption{Risk classification of In-Vitro Diagnostic (IVD) medical devices according to HSA, Singapore.}
 \label{tab:risk_class_ivd_medical}
\begin{tabular}{c|c|c} \hline
\rowcolor{gray!30} \textbf{CLASS} & \textbf{RISK LEVEL} & \textbf{EXAMPLES} \\ \hline
\multirow{2}{*}{A (IVD)} & Low Individual Risk & Specimen collection tubes \\       & Low Public Health Risk & General culture media \\\hline
\multirow{2}{*}{B (IVD)} & Moderate Individual Risk  & Pregnancy tests, Anti-Nuclear  \\ 
      & Low Public Health Risk & Antibody tests, Urine test strips\\ \hline
\multirow{2}{*}{C (IVD)} & High Individual Risk & Blood glucose tests, HLA typing tests,  \\ 
      & Moderate Public Health Risk & PSA screening tests, Rubella tests \\\hline
\multirow{2}{*}{D (IVD)} & High Individual Risk & Screening for HIV,  \\ 
    & High Public Health Risk  & ABO blood grouping tests\\\hline
\end{tabular}
\end{table}


Similarly, the UK Medicines and Healthcare products Regulatory Agency (MHRA) classifies IVD devices into classes A, B, C, and D, where class A represents the lowest risk and typically does not require formal approval. For general (non-IVD) medical devices, the UK follows a classification system comprising classes I, IIa, IIb, and III, with class III reserved for the highest-risk devices.

In the United States, the Food and Drug Administration (FDA) uses a three-tier classification: Class I, II and III. Class I devices \textemdash such as toothbrushes or adhesive bandages \textemdash are considered low-risk and require only registration and listing, without the need for FDA clearance or approval. Class II devices typically require approval from the FDA through the pre-market notification process (510(k)), while Class III devices \textemdash such as pacemakers and other implantable devices \textemdash must undergo rigorous pre-market approval involving clinical trials and FDA review.

\subsection{Regulatory Flows Integrated with Cybersecurity}
In recent years, growing concerns have emerged regarding the cybersecurity risks associated with medical devices. One of the earliest signals of this shift was the recall of a medical device in the United States due to cybersecurity vulnerabilities~\cite{insulin_security_recall}. In Singapore, the identification of a critical Bluetooth vulnerability~\cite{sweyntooth} prompted the Health Sciences Authority (HSA) to issue a public warning~\cite{swyentooth_hsa}, highlighting the seriousness of such threats. As a result, cybersecurity considerations have progressively been integrated into medical device regulatory frameworks~\cite{US_FDA_Cyber, cybersecurity_best_practices_hsa, mdcg_europe_cybersecurity}. In particular, Singapore's Cyber Security Agency (CSA) has introduced a cybersecurity labeling scheme that classifies devices into security levels, each with specific requirements for inclusion. Table~\ref{tab:csa_md_label} presents the classification levels and their corresponding criteria.

\begin{table}[hbt]
\centering
\caption{Cybersecurity Requirements by Level}
\label{tab:csa_md_label}
\begin{tabular}{l|c}
\hline
\rowcolor{gray!30} \textbf{Level} & \textbf{Requirements} \\
\hline
Level 1 & Meets baseline cybersecurity requirements. \\
\hline
Level 2 & Meets enhanced cybersecurity requirements. \\
\hline
\multirow{3}{*}{Level 3} & Meets enhanced cybersecurity requirements. \\
        & Will be required to pass independent \\
        & third-party software binary analysis and penetration testing. \\
\hline
\multirow{3}{*}{Level 4} & Meets enhanced cybersecurity requirements. \\
        & Will be required to pass independent \\
        & third-party software binary analysis and security evaluation. \\
\hline
\end{tabular}
\end{table}

In general, the commercialization of medical devices \textemdash including those within the IoMT ecosystem \textemdash follows a well-defined regulatory pathway. In Singapore, the national guideline for best cybersecurity practices~\cite{cybersecurity_best_practices_hsa} outlines a comprehensive strategy for security testing. This includes procedures such as \textit{vulnerability assessment}, \textit{penetration testing}, \textit{security audit}, and \textit{security configuration review}, among others. These pre-market evaluations must be integrated with a clear post-market cybersecurity plan that encompasses vigilance, coordinated vulnerability disclosure, patch management and updates, system recovery procedures, and structured information sharing. In addition, the guidelines require the establishment of a contractual agreement between the medical device manufacturer and the healthcare service provider. This agreement must include a \textbf{Product Life Cycle Document (PLCD)}, which details critical device information such as the operating system in use, security scanning capabilities, a Software Bill of Materials (SBOM) to identify all software components, and a list of required ports and services necessary for proper device functionality. Importantly, when manufacturers opt to apply for a cybersecurity labeling scheme, these technical details must be included in the regulatory submission. Consequently, manufacturers are expected to implement the corresponding countermeasures outlined in Section~\ref{sec:translation_countermeasures} of this document.

In the United States, the cybersecurity guidelines for medical devices are issued by the Food and Drug Administration (FDA)~\cite{US_FDA_Cyber}. A detailed cybersecurity assessment report, along with its implications for patient safety, must be included as part of the pre-market approval and FDA clearance documentation. The guidelines strongly recommend managing cybersecurity throughout the Total Product Life Cycle (TPLC), acknowledging that cybersecurity is a dynamic and evolving challenge. The FDA's cybersecurity assessment begins with defining security objectives such as confidentiality, integrity, authentication, authorization, availability, and timely patchability. The guide makes a clear distinction between \textit{safety risk management} and \textit{security risk management}. While safety risk management focuses on patient harm, security risk management centers on identifying threats and mitigating exploitable vulnerabilities. These aspects are further elaborated in the FDA post-market cybersecurity management guidelines. In particular, the scope of traditional safety risk management (as defined in ISO 14971) is extended by the Association for the Advancement of Medical Instrumentation (AAMI) through the Technical Information Report TIR57:2016 (R2023), which provides additional direction for incorporating cybersecurity into medical device risk analysis. Although the FDA currently does not classify devices by cybersecurity risk level, it requires the inclusion of detailed labeling and a cybersecurity management plan as part of regulatory submissions. Such labeling may include a Software Bill of Materials (SBOM), 
security scanning capabilities (e.g., Intrusion Detection Systems), backup and restoration procedures, and verified mechanisms for downloading manufacturer-authenticated software updates, among other details.


\begin{figure}[hbt]
    \centering
    \includegraphics[width=\linewidth]{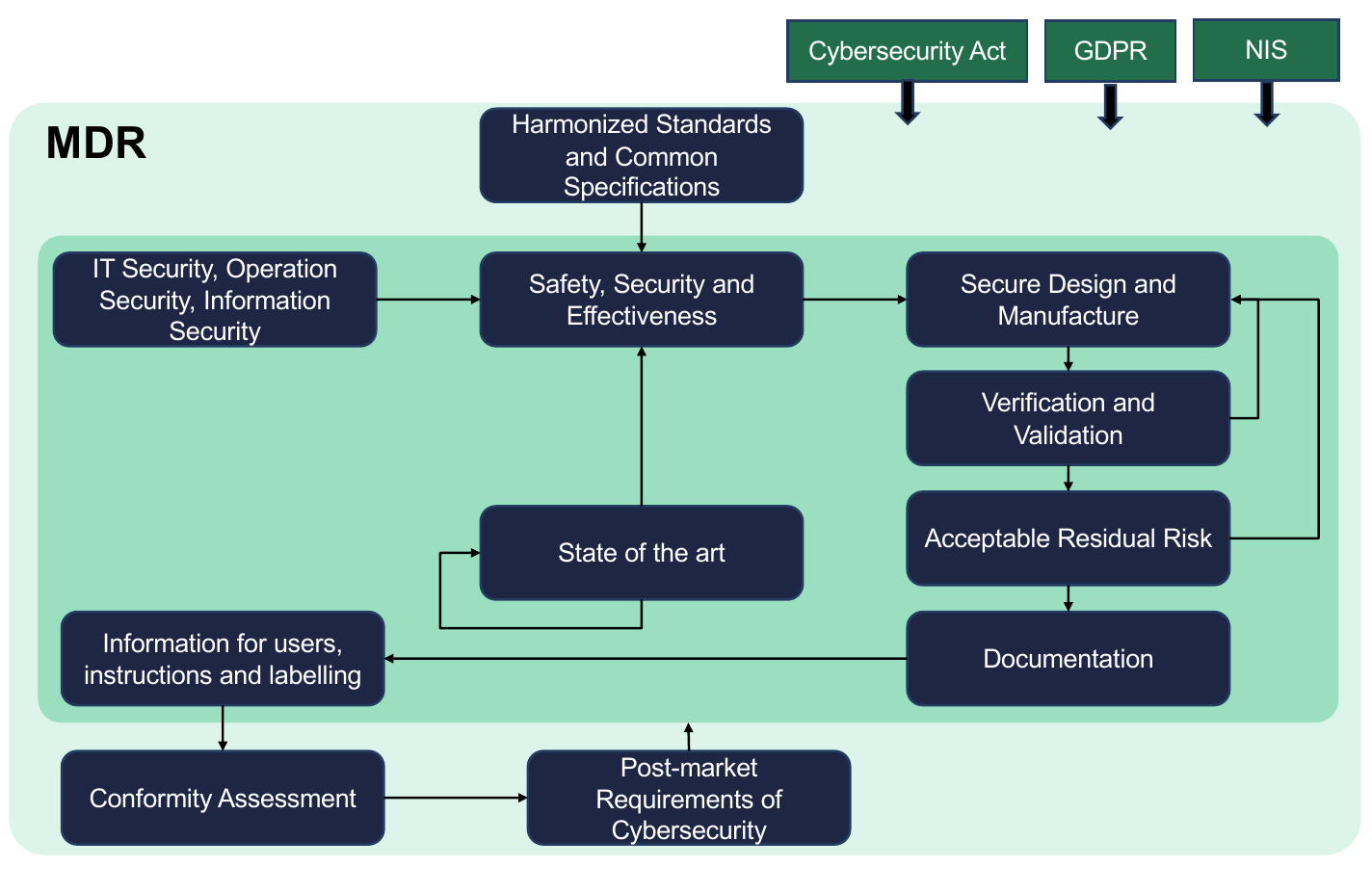}
    \caption{Cybersecurity Requirements in the EU Medical Device Regulations}
    \label{fig:eu_mdr_flow}
\end{figure}


The cybersecurity framework under the European Union Medical Device Regulation (EU MDR) is schematically illustrated in Fig.~\ref{fig:eu_mdr_flow}. In addition to MDR, cybersecurity of medical devices must also comply with related regulations, including the EU Cybersecurity Act, the General Data Protection Regulation (GDPR), and the Directive on Security of Network and Information Systems (NIS Directive). Annex II of the EU MDR cybersecurity guidance~\cite{mdcg_europe_cybersecurity} provides several examples illustrating how a security risk can translate into a safety risk. In response, the guidelines distinguish between two types of controls: \textit{security controls} and \textit{safety controls}.
\begin{itemize}
    \item Security controls aim to prevent vulnerabilities from being exploited.
    \item Safety controls are designed to prevent an exploited vulnerability from resulting in a safety-related hazard.
\end{itemize}


Despite strong enforcement by regulatory bodies, current certification efforts often focus on individual medical devices, for which manufacturers can obtain approval independently. However, the growing adoption of connected medical devices~\cite{ghubaish2020recent}, the rise of sophisticated attack vectors such as side-channel attacks, and persistent vulnerabilities in widely used network protocols~\cite{sweyntooth} highlight that significant gaps remain \textemdash both for security practitioners and standardization bodies. Addressing these challenges requires coordinated global efforts. Initiatives such as those led by the International Medical Device Regulators Forum (IMDRF) play a critical role in promoting international collaboration and harmonization of cybersecurity standards.

\section{Conclusion and Future Roadmap}
In summary, we conducted an extensive survey of state-of-the-art cyberattacks targeting networked medical devices, uncovering a diverse range of attack surfaces and vulnerabilities across different layers of the IoMT architecture. To systematically classify these threats, we proposed a structured attack taxonomy that categorizes vulnerabilities within distinct layers of the IoMT ecosystem, including their respective communication channels. This taxonomy not only provides a comprehensive understanding of how cyber threats propagate through IoMT systems, but also serves as a foundation for designing targeted security measures. 

Building on this taxonomy, we introduced a layer-wise security assessment framework designed to assist security engineers, network administrators, and medical device manufacturers in identifying, evaluating, and mitigating vulnerabilities at each architectural level. This structured methodology enables a granular and risk-informed approach to IoMT security, promoting more robust and resilient system designs.

As cyber threats in the healthcare domain continue to evolve, future research must focus on developing proactive defense mechanisms, including AI-driven threat detection, cryptographic advancements, and secure-by-design medical device architectures. In addition, regulatory frameworks and industry-wide collaboration will be essential to establishing robust security standards that address emerging threats while ensuring the seamless functionality of IoMT systems. 

Importantly, securing IoMT requires a multi-layered, proactive approach that integrates technical safeguards, policy enforcement, and continuous monitoring. The taxonomy and methodology presented here aim to serve as a foundational tool for advancing cybersecurity in modern healthcare, contributing to the realization of trustworthy and resilient IoMT ecosystems.

IoMT security also necessitates a thorough examination of adjacent domains that fall beyond the scope of this manuscript. Notably, this includes the development of a security-driven risk management framework, which plays a vital role in meeting regulatory requirements. Furthermore, there is a pressing need to distinguish between safety-related and security-related controls and to establish a clear link between the two. These aspects warrant deeper investigation, particularly in light of the rapidly evolving cybersecurity landscape.

\bibliographystyle{ieeetr}

\end{document}